\pdfoutput=1
\newcommand*{\ATLASLATEXPATH}{}
\documentclass[PAPER, atlasdraft=false, texlive=2016, cernpreprint, UKenglish,LANGEDIT=false]{\ATLASLATEXPATH atlasdoc}
\PreprintIdNumber{CERN-EP-2018-291}
\usepackage[subfigure=true]{\ATLASLATEXPATH atlaspackage}
\usepackage{\ATLASLATEXPATH atlasbiblatex}
 
\usepackage{\ATLASLATEXPATH atlasphysics}
 
\graphicspath{{logos/}{figures/}}
 
\usepackage{ANA-BPHY-2018-09-PAPER-defs}

\AtlasTitle{Study of the rare decays of $B^0_s$ and $B^0$ mesons into muon pairs
using data collected during 2015 and 2016 with the ATLAS detector}
 
\AtlasVersion{2.9}
\AtlasAbstract{
A study of the decays $B^0_s \to \mu^+\mu^-$ and $B^0 \to \mu^+\mu^-$ has been performed using
$26.3$~fb$^{-1}$ of $13$ $\text{Te\kern -0.1em V}$ LHC proton--proton collision data collected
with the ATLAS detector in 2015 and 2016.
Since the detector resolution in $\mu^+\mu^-$ invariant mass is comparable to the
$B_{s}^{0}$--$B^0$ mass difference, a single fit determines the signal yields
for both decay modes.
This results in a measurement of the branching fraction
${\cal B}(B^0_s \to \mu^+\mu^-) = \left( 3.2^{+1.1}_{-1.0} \right) \times 10^{-9}$
and an upper limit
${\cal B}(B^0 \to \mu^+\mu^-) < 4.3 \times 10^{-10}$ at $95\%$ confidence level.
The result is combined with the Run~1 ATLAS result, yielding ${\cal B}(B^0_s \to \mu^+\mu^-) =  \left( 2.8^{+0.8}_{-0.7} \right) \times 10^{-9}$ and
${\cal B}(B^0 \to \mu^+\mu^-) < 2.1 \times 10^{-10}$ at 95\%
confidence level. The combined result is consistent with the Standard
Model prediction within 2.4 standard deviations in the ${\cal B}(B^0
\to \mu^+\mu^-)$-${\cal B}(B^0_s \to \mu^+\mu^-)$ plane.
}
 
\author{The ATLAS Collaboration}
 
\AtlasRefCode{BPHY-2018-09}

\AtlasJournal{Journal of High Energy Physics}
\AtlasJournalRef{JHEP 04 (2019) 098}
\AtlasDOI{10.1007/JHEP04(2019)098}
 
\AtlasCoverSupportingNote{ANA-BPHY-2018-09-INT1}{https://cds.cern.ch/record/2311593}
 
\AtlasCoverCommentsDeadline{31 Oct 2018}
 
\AtlasCoverAnalysisTeam{A. Cerri, I. Ibragimov, A. Grummer, F. Tresoldi, U. de Sanctis, A. Campoverde, W. Walkowiak, P. Buchholz, S. Seidel, S.Yu. Sivoklokov}
 
\AtlasCoverEdBoardMember{Julie~Kirk, Bill~Murray~(chair), Sandro~Palestini}

\AtlasCoverEgroupEditors{atlas-BPHY-2018-09-editors@cern.ch}
 
\AtlasCoverEgroupEdBoard{atlas-BPHY-2018-09-editorial-board@cern.ch}

\hypersetup{pdftitle={ATLAS document},pdfauthor={The ATLAS Collaboration}}
 
\begin{document}
 
\maketitle

\section{Introduction}
\label{sec:introduction}
 
Flavour-changing neutral-current processes are highly suppressed
in the Standard Model (SM). 
The branching fractions of the decays \Bmumu\
are, in addition, helicity suppressed 
in the SM, and are predicted to be
$\BR(\Bsmumu) = (3.65 \pm 0.23) \times 10^{-9}$ and
$\BR(\Bdmumu) = (1.06 \pm 0.09) \times 10^{-10}$~\cite{Bobeth:2013uxa}.
The smallness and precision of these predicted branching fractions provide a favourable
environment for observing contributions
from new physics: significant deviations from SM predictions could arise in models involving
non-SM heavy particles, such as those predicted in the
Minimal Supersymmetric Standard Model~\cite{Huang:1998vb,
Hamzaoui:1998nu,Choudhury:1998ze,Babu:1999hn,Choudhury:2005rz} and
in extensions such as Minimal Flavour Violation~\cite{DAmbrosio:2002ex, Buras:2003td},
Two-Higgs-Doublet Models~\cite{Choudhury:2005rz},
and others~\cite{Davidson:2010uu,Guadagnoli:2013mru}.
The CMS and LHCb collaborations reported the observation of
\Bsmumu~\cite{Chatrchyan:2013bka,Aaij:2013aka} and evidence of
\Bdmumu\ with combined values
$\BR(\Bsmumu) = \left(2.8^{+0.7}_{-0.6}\right) \times 10^{-9}$ and
$\BR(\Bdmumu) = \left(3.9^{+1.6}_{-1.4}\right) \times 10^{-10}$~\cite{CMS:2014xfa}.
The LHCb Collaboration updated its Run~1 result with part of the data collected in Run~2,
measuring $\BR(\Bsmumu) = \left(3.0\pm 0.6^{+0.3}_{-0.2}\right) \times 10^{-9}$ (where the first uncertainty is statistical and the second
systematic) and determining $\BR(\Bdmumu) < 3.4 \times 10^{-10}$ at $95\%$ confidence level (CL)~\cite{Aaij:2017vad}.
ATLAS has measured, with the Run~1 dataset~\cite{Aaboud:2016ire}, $\BR(\Bsmumu) = \left(0.9^{+1.1}_{-0.8}\right) \times 10^{-9}$, while setting an upper limit on $\Bdmumu$ of $4.2 \times 10^{-10}$ at $95\%$ CL.
 
This paper reports the result of a search for \Bsmumu\ and \Bdmumu\
decays performed using $pp$ collision data corresponding to an effective integrated
luminosity of 26.3~fb$^{-1}$, collected at $13$~TeV centre-of-mass energy during the first two years of the LHC
Run~2 data-taking period using the ATLAS detector.
The analysis strategy follows the approach employed in the previous ATLAS measurement~\cite{Aaboud:2016ire},
but uses data collected with one rather than three separate sets of trigger thresholds,
applies standard ATLAS muon selection criteria (Section~\ref{sec:sel}) rather than a dedicated muon multivariate discriminant,
and employs an improved statistical treatment of the result (Section~\ref{sec:sigfit}).
 
\label{sec:outline}
 
The notation used throughout the paper refers to the combination of processes and their
charge-conjugates, unless otherwise specified.
The \Bsmumu\ and \Bdmumu\ branching fractions are measured relative
to the reference decay mode $B^+\to J/\psi (\to \mu^+\mu^-)K^+$
which is
abundant and has a well-measured branching fraction $\BR (\BpKpJpsi) \times \BR (J/\psi \to \mumu)$.
The \Bdmumu\ (\Bsmumu) branching fraction can be extracted as:
\begin{eqnarray}
\BR (B^0_{(s)}\! \to \!\mu^+ \mu^-) &=& \frac{N_{d(s)}}{\varepsilon_{\mu^+\mu^-}}
\times \left[\BR (\BpKpJpsi) \times \BR (J/\psi \to \mumu)\right]
\frac{\varepsilon_{J/\psi K^+}}{N_{J/\psi K^+}}
\times \frac{f_{u}}{f_{d(s)}}\nonumber \\
&=& N_{d(s)}\frac{\BR (\BpKpJpsi) \times \BR (J/\psi \to \mumu)}{\mathcal{D}_{\mathrm{ref}}}\times \frac{f_{u}}{f_{d(s)}}\, , 
\label{eq:BRFormula}
\end{eqnarray}
where $N_d$ ($N_s$) is the \Bdmumu\ (\Bsmumu) signal yield,
$N_{J/\psi K^+}$ is the \BpKpJpsi\ reference channel yield,
$\varepsilon_{\mu^+\mu^-}$ and $\varepsilon_{J/\psi K^+}$
are the corresponding values of acceptance times efficiency (measured in fiducial
regions
defined in Section~\ref{sec:eff}),
and $f_u/f_d$  ($f_u/f_s$) is the ratio of the hadronisation probabilities of a
$b$-quark into \Bp\ and \Bz\ (\Bs).
In the quantity $\mathcal{D}_{\mathrm{ref}} =  N_{J/\psi K^+} \times (\varepsilon_{\mu^+\mu^-}/\varepsilon_{J/\psi K^+})$, the $\varepsilon$ ratio takes into account relative differences in efficiencies, integrated luminosities and
the trigger selections used for the signal and the reference modes.
Signal and reference channel events are selected with similar dimuon triggers.
One half of the reference channel sample is used to determine the normalisation and the other half is used
to tune the kinematic distributions of simulated events.
 
The event selection uses variables related to the $B$ candidate
decay time, thus introducing a dependence of the efficiency on the signal lifetime.
The relation between the measured branching fraction and the corresponding value
at production is established assuming the decay time distribution predicted in the SM,
where the decay occurs mainly through the heavy eigenstate
$B^0_{(s),\mathrm{H}}$ of the $B^0_{(s)}$--$\overline{B^0_{(s)}}$ system.
Some models of new physics~\cite{DeBruyn:2012wk, Buras:2013uqa}
predict modifications to the decay time distribution of \Bsmumu and a comparison
with the experimental result requires a correction to the ratio of the
time-integrated efficiencies entering $\mathcal{D}_{\mathrm{ref}}$.
 
The ATLAS inner tracking system, muon spectrometer and, for efficient identification
of muons, also the calorimeters, are used to reconstruct
and select the event candidates. Details of the detector, trigger, data sets,
and preliminary selection criteria are discussed in Sections~\ref{sec:data} and~\ref{sec:sel}.
A blind analysis was performed in which data in the dimuon invariant mass range
from $5166$ to $5526$~\MeV\ were removed until the procedures
for event selection and the details of signal yield extraction
were completely defined.
Section~\ref{sec:bkg} introduces the three main categories of background.
Section~\ref{sec:fakes} describes the strategy used to reduce the probability of hadron misidentification.
The final sample of candidates is selected using a multivariate classifier,
designed to enhance the signal relative to the dominant dimuon background component,
as discussed in Section~\ref{sec:contBDT}.
Checks on the distributions of the variables used in the
multivariate classifier are summarised in Section~\ref{sec:comp}.
They are based on the comparison of data and simulation
for dimuon events, for \BpKpJpsi\ candidates and for events selected
as $\BsJpsiPhi \to \mu^+\mu^- K^+K^-$, which provide an additional
validation of the procedures used in the analysis.
Section~\ref{sec:bplus} details the fit procedure used to extract the yield of
\BpKpJpsi\ events. The determination of the
ratio of efficiencies in the signal and the reference channels
is presented in Section~\ref{sec:eff}.
Section~\ref{sec:sigfit} describes the extraction of the signal yield,
obtained with an unbinned maximum-likelihood fit performed on the dimuon
invariant mass distribution. In this fit, events are separated into classifier
intervals to maximise the fit sensitivity.
The results for the branching fractions \BR(\Bsmumu)\ and \BR(\Bdmumu) are
reported in Section~\ref{sec:br}
and combined with the full Run~1 results
in Section~\ref{sec:combination}.
\section{ATLAS detector, data and simulation samples}\label{sec:data}
 
The ATLAS detector\footnote{ATLAS uses a right-handed coordinate system
with its origin at the nominal interaction point. The $z$-axis is along
the beam pipe, the $x$-axis points to the centre of the LHC ring and
the $y$-axis points upward. Cylindrical coordinates $(r,\phi)$ are used
in the transverse plane, $r$ being the distance from the origin and
$\phi$ being the azimuthal angle around the beam pipe. The pseudorapidity
$\eta$ is defined as $\eta = - \ln [\tan (\theta/2)]$ where
$\theta$ is the polar angle.}
consists of three main components: an inner detector (ID) tracking system
immersed in a $2$~T axial magnetic field, surrounded by electromagnetic and hadronic
calorimeters and by the muon spectrometer (MS). A full
description can be found in Ref.~\cite{AtlasDet}, complemented by Ref.~\cite{IBLpaper}
for details about the new innermost silicon pixel layer that was installed for Run~2.
 
This analysis is based on the Run~2 data recorded in 2015
and 2016 from $pp$ collisions at the LHC at
$\sqrt{s}=13$~\TeV.
Data used in the analysis were recorded during stable LHC beam periods.
Data quality requirements were imposed, notably on the
performance of the MS, ID and calorimeter systems.
The total integrated luminosity collected by ATLAS in this period is
$36.2$~fb$^{-1}$ with an uncertainty of 2.1\%.
These values are determined using a methodology similar to that detailed in Ref.~\cite{DAPR-2013-01},
based on calibration of the luminosity scale using $x$--$y$ beam-separation scans,
and use the LUCID-2 detector~\cite{LUCID2} for the baseline luminosity measurement.
The total effective integrated luminosity used in this analysis - accounting for trigger prescales -
amounts to $26.3$~fb$^{-1}$ for the signal and $15.1$~fb$^{-1}$ for the reference channel.

Samples of simulated Monte Carlo (MC) events are used for training and validation
of the multivariate analyses, for the determination of the efficiency
ratios, and for 
developing the procedure used to determine the signal.
Exclusive MC samples
were produced for the signal channels $\Bsmumu$ and $\Bdmumu$, the
reference channel \BpKpJpsi\ (\Jpsimumu),
and the control channel $\BsJpsiPhi$ (\Jpsimumu, $\Phikk$).
In addition, background studies employ MC samples
of inclusive $b\bar{b}\to\mumu X$ decays, exclusive
samples of $\Bs \to K^- \mu^+ \nu$, $B^0 \to \pi^-\mu^+ \nu$, $\Lambda_b \to p \mu^- \overline{\nu}$,
$\Bhh$ decays with $h^{(\prime)}$ being a charged pion or kaon, and inclusive decays $B \to J/\psi X$ as
well as the exclusive \BpPipJpsi\ decay.
 
Most of the dimuon candidates in the data sample originate from the 
decays of hadrons produced in the hadronisation of $b\bar{b}$ pairs.
The inclusive $b\bar{b}\to\mumu X$ MC sample used to describe this background
requires the presence of two muons in the final state,
with both muons originating from the $b\bar{b}$ decay chain.
The size of this sample is equivalent to roughly
three times the integrated luminosity of the data.
 
The MC samples were generated with \Pythia~8~\cite{Sjostrand:2007gs}.
The ATLAS detector and its response were simulated using \Geant~\cite{GEANT4,Aad:2010ah}.
Additional $pp$ interactions in the same and nearby bunch crossings
(pile-up) are included in the simulation.
Muon reconstruction and triggering efficiencies are corrected in the simulated samples
using data-driven scale factors. The scale factors for the trigger efficiencies
are obtained by comparing data and simulation efficiencies determined
with a tag-and-probe method on inclusive prompt and non-prompt $J/\psi$ candidates.
This procedure yields scale factors as a function of
the muon transverse momentum and pseudorapidity, which are applied throughout the
analysis~\cite{tag-n-probe-Run2}.
Reconstruction and selection efficiencies are obtained from simulation and similarly
corrected according to data-driven comparisons.
In addition to these efficiency corrections, simulated events are reweighted
to reproduce the pile-up multiplicity observed in data, and
according to the equivalent integrated luminosity associated with each trigger
selection.

Using the iterative reweighting method described in Ref.~\cite{bsmumuv1},
the simulated samples of the exclusive decays considered are adjusted with
two-dimensional data-driven weights (DDW) to correct for the
differences between simulation and data observed in the $B$ meson transverse momentum and
pseudorapidity distributions.
DDW obtained from \BpKpJpsi\ decays are used to correct the
simulation samples in the signal and reference channels.
DDW obtained from the \BsJpsiPhi\ control channel are found to
agree with those from \BpKpJpsi,
showing that the same corrections are applicable to \Bs and \Bz decays.
 
Residual differences between data and simulation studied in the \BpKpJpsi\ and \BsJpsiPhi\ signals are treated as sources
of systematic uncertainty in the evaluation of the signal efficiency, as discussed in Section~\ref{sec:eff}.
The only exception to this treatment is the $B$ meson isolation ($I_{0.7}$ in Section~\ref{sec:contBDT} and Table~\ref{tab:15vars}),
where residual differences are used to reweight the signal MC events and the corresponding
uncertainties are propagated to account for residual systematic uncertainty effects.
 
Similarly to the exclusive decays, the kinematic distributions of the
inclusive $b\bar{b}\to\mumu X$ MC sample
are reweighted with corrections obtained from the dimuon invariant mass sidebands in data.

\section{Data selection}\label{sec:sel}
 
For data collected during the LHC Run~2, the ATLAS detector uses
a two-level trigger system, consisting of a hardware-based first-level trigger and
a software-based high-level trigger. A first-level dimuon trigger~\cite{Aaboud:2016leb} 
selects events requiring that one muon
has $\pt > 4$~\GeV\ and the other has $\pt > 6$~\GeV.
A full track reconstruction of the muon candidates is performed by the
high-level trigger, where an additional loose selection is imposed on the dimuon invariant
mass $m_{\mu\mu}$, accepting candidates in the range 4 \GeV\ to 8.5 \GeV.
Due to the increased pile-up in 2016 data, an additional selection was added at this trigger stage,
requiring the vector from the primary vertex to the dimuon vertex to have a positive component ($L_{xy}$)
along the dimuon's transverse momentum direction. The effect of this selection is accounted for in the analysis but
has no consequence since stricter requirements are applied in the full event selection (see Section~\ref{sec:contBDT}).
 
The signal channel, the reference channel \BpKpJpsi\ and the control channel $\Bs \to J/\psi \phi$
were selected with trigger prescale factors that vary during the data-taking period.
In the 36.2~fb$^{-1}$ of data analysed, the prescaling of the trigger approximately averages to
a reduction by a factor 1.4, giving an effective integrated luminosity for the signal sample
of 26.3~fb$^{-1}$, while for the reference and control channels 15.1~fb$^{-1}$ were collected due to an effective
prescale of 2.4.
These effects are taken into account in the extraction of the signal branching fraction,
through the $\varepsilon$ factors
in Eq.~(\ref{eq:BRFormula}).
 
Using information from the full offline reconstruction,
a preliminary selection is performed on candidates
for $\Bmumu$, $\BpKpJpsi \to \mumu K^+$ and \Bs$ \to J/\psi \phi \to \mumu K^+ K^-$ decays.
In the ID system, muon candidates are required to have at least one measured hit in the pixel detector and
two measured hits
in the semiconductor tracker.
They are also required to be reconstructed in the MS, and to have $|\eta| < 2.5$.
The offline muon pair must pass the $\pt>4$~\GeV~and $\pt>6$~\GeV~requirements imposed by the trigger.
Furthermore, the muon candidates are required to fulfil \textit{tight} muon quality criteria~\cite{PERF-2015-10};
this requirement is
relaxed to \textit{loose} for the hadron misidentification studies in Section~\ref{sec:fakes}.
Kaon candidates must satisfy similar requirements in the ID, except for a looser requirement of $\pt>1$~\GeV.
 
The computed $B$ meson properties are based on a decay vertex fitted to two,
three or four tracks, depending on the decay process to be reconstructed.
The $B$ candidates are required to have a $\chi^2$ per degree of freedom below 6 for the
fit to the $B$ vertex, and below 10 for the fit to the $J/\psi \to \mumu$ vertex.
The selections $2915<m(\mumu)<3275$~\MeV\ and
$1005 < m(K^+K^-) < 1035$~\MeV\ are  applied to the
$J/\psi \to \mumu$ and the $\phi \to K^+K^-$ vertices, respectively.
In the fits to the \BpKpJpsi\ and \BsJpsiPhi\ channels, the reconstructed dimuon mass is
constrained to the world average $J/\psi$ mass~\cite{PDG2018}.
 
Reconstructed $B$ candidates are retained if they satisfy $\pt^B>8.0$~\GeV\
and $|\eta^B|<2.5$.
The invariant mass of each $B$ candidate is calculated using muon trajectories measured
by combining the information from the ID and MS to improve upon the mass resolution
obtained from ID information only~\cite{Aad:2014rra}.
 
The invariant mass range considered for the $\Bmumu$ decay starts at $4766$~\MeV\ and is $1200$~\MeV\ wide.
Within this range a $360$-\MeV-wide signal region is defined, starting at $5166$~\MeV.
The remainder of the range defines the upper and lower mass sidebands of the analysis.
 
For the reference and control channels, the mass range considered is
$4930$--$5630$ ($5050$--$5650$)~\MeV\ for \BpKpJpsi\ (\BsJpsiPhi),
where $5180$--$5380$ ($5297$--$5437$)~\MeV\ is the peak
region and higher and lower mass ranges comprise the mass sidebands
used for background subtraction.
 
The coordinates of primary vertices (PV) are obtained from charged-particle tracks not used in the decay
vertices, and that are constrained to the luminous region of the colliding beams in the transverse plane.
The matching of a $B$ candidate to a PV is made by extrapolating the candidate trajectory
to the point of closest approach to the beam axis, and choosing the PV with
the smallest distance along $z$. Simulation shows that this method
matches the correct vertex with a probability above $99\%$ for all relevant pile-up conditions.
 
To reduce the large background in the \Bmumu\ channel
before applying the final selection based on multivariate classifiers,
a loose collinearity
requirement is applied between the momentum of the $B$ candidate
($\overrightarrow{p}^B$) and the 
vector from the PV to the decay vertex ($\overrightarrow{\Delta x}$).
The absolute value of the azimuthal angle $\alpha_{2\mathrm{D}}$ between these two vectors is required to be smaller than
1.0 radians.  The combination $\Delta R_{\mathrm{flight}} = \sqrt{\alpha_{2\mathrm{D}}{}^{2} + (\Delta\eta)^{2}}$, where $\Delta\eta$ is the difference in pseudorapidity,
is required to satisfy $\Delta R_{\mathrm{flight}}<1.5$.
 
After the preliminary selection, approximately $3.5\times 10^6$
candidates are found in the $\Bmumu$ fit region, with about $1.0 \times 10^6$
falling in the blinded range [5166, 5526] \MeV .
 
\section{Background composition}
\label{sec:bkg}
 
The background to the \Bmumu\ signal originates from three main sources:
\begin{description}
\setlength{\itemsep}{0pt}
\setlength{\parskip}{0pt}
\item[] \textit{Continuum} background, the dominant combinatorial component, which consists
of muons originating from uncorrelated hadron decays and is characterised
by a weak dependence on the dimuon invariant mass;
\item[] \textit{Partially reconstructed}  decays,
where one or more of the
final-state particles ($X$) in a $b$-hadron decay
is not reconstructed, causing these candidates to accumulate
in the low dimuon invariant mass sideband
(this background includes a significant contribution from semileptonic decays where
one of the  muons is a misidentified hadron, discussed below);
\item[] \textit{Peaking} background, due to \Bhh\ decays,
with both hadrons misidentified as muons.
\end{description}
 
The continuum background consists mainly of
muons produced independently in the fragmentation and decay chains of a
$b$-quark and a $\bar{b}$-quark.
It is studied in the signal mass sidebands, and it is found to be
well described by the inclusive $b\bar{b}\to\mumu X$ MC sample.

The partially reconstructed decays consist of several topologies:
(a) {\em same-side} combinatorial background from decay cascades
($b \to c \mu^-\overline{\nu} \to  s(d) \mumu\nu\overline{\nu}$);
(b) {\em same-vertex} background from $B$ decays containing a muon pair
(e.g. $B^0 \to K^{*0}\mumu$ or $B \to J/\psi \mu^+ X \to \mumu\mu^+ X$);
(c) $B_c^+$ decays (e.g. $B_c^+ \to J/\psi \mu^+ \nu \to \mumu\mu^+\nu$);
(d) semileptonic $b$-hadron decays where a final-state hadron is
misidentified as a muon.
The remainder of this paper implicitly excludes categories (c) and (d) when referring to
partially reconstructed or $b\to\mumu X$ decays, since these categories are treated separately.
 
The $b\bar{b}\to\mumu X$ MC sample is used to investigate the background composition
after the analysis selection.
All backgrounds in this sample have a dimuon invariant mass distribution mainly below the mass range considered in this analysis, with a high-mass tail extending through the signal region.
The simulation does not contemplate sources other than muons from
$b\bar{b}$ decays: $c\bar{c}$ and prompt contributions are not
included\footnote{These sources are suppressed by the final
analysis selections introduced in Section
\ref{sec:contBDT}. Potential residual contributions are found to be
consistent with the continuum background models used in the final fit
(Section \ref{sub:sec:model}), where systematic uncertainties for
background model inconsistencies between data and MC are taken into account (Section \ref{sub:sec:fitsyst}).}.
All possible origins of two muons in the $b\bar{b}$ decay tree are, however, analysed,
after classification into the mutually exclusive continuum and partially reconstructed categories described
above.
This sample is
used only to identify suitable functional models for the corresponding background
components, and as a benchmark for these models. No shape or normalisation constraints are derived from
this simulation. This makes the analysis largely insensitive to mismatches between background simulation and data.
 
\begin{figure}[!htb]
\begin{center}
\subfigure[]{
\hspace*{-0.5cm}
\includegraphics[width=0.52\textwidth]{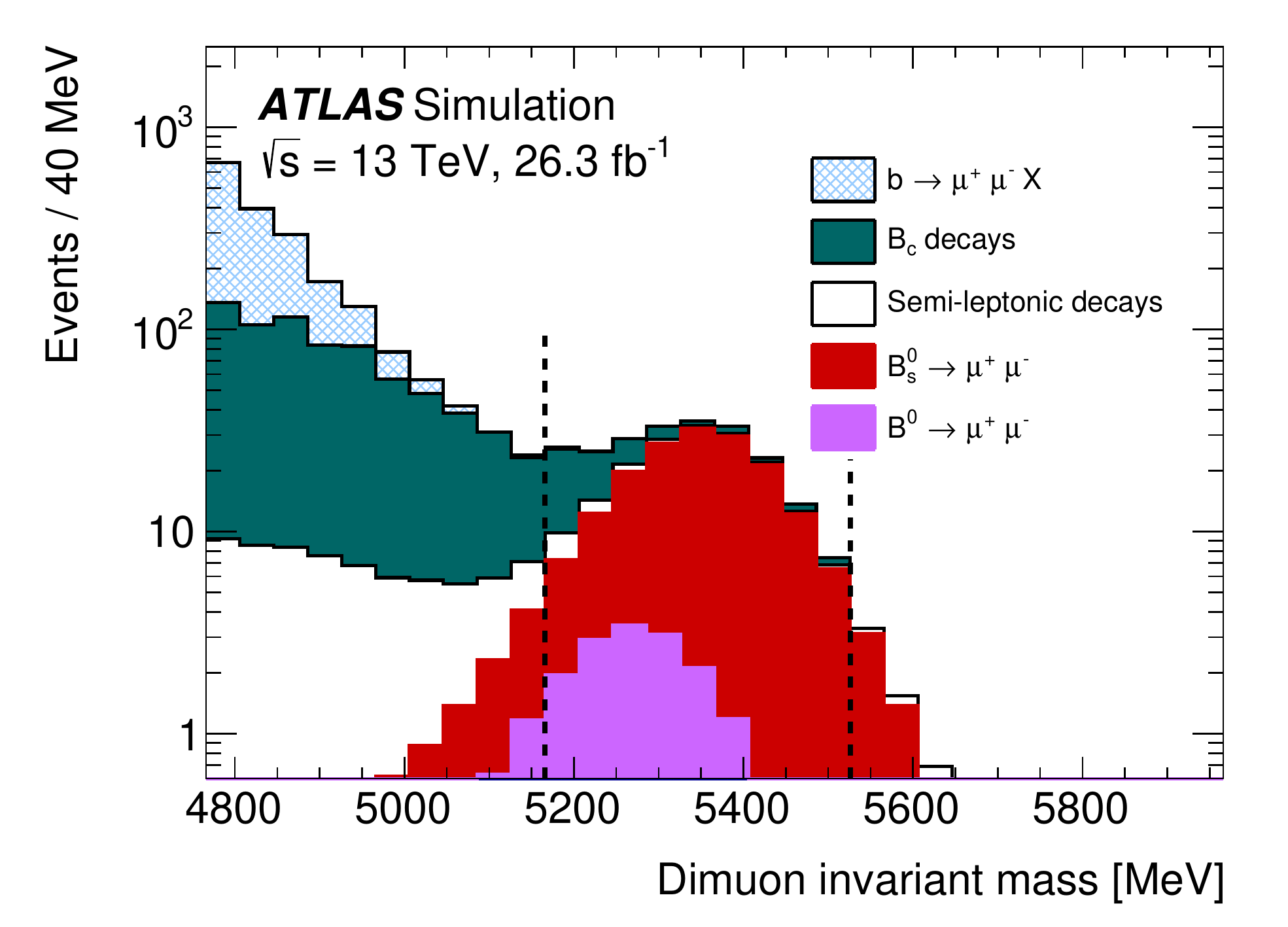}
}
\subfigure[]{
\hspace*{-0.5cm}
\includegraphics[width=0.52\textwidth]{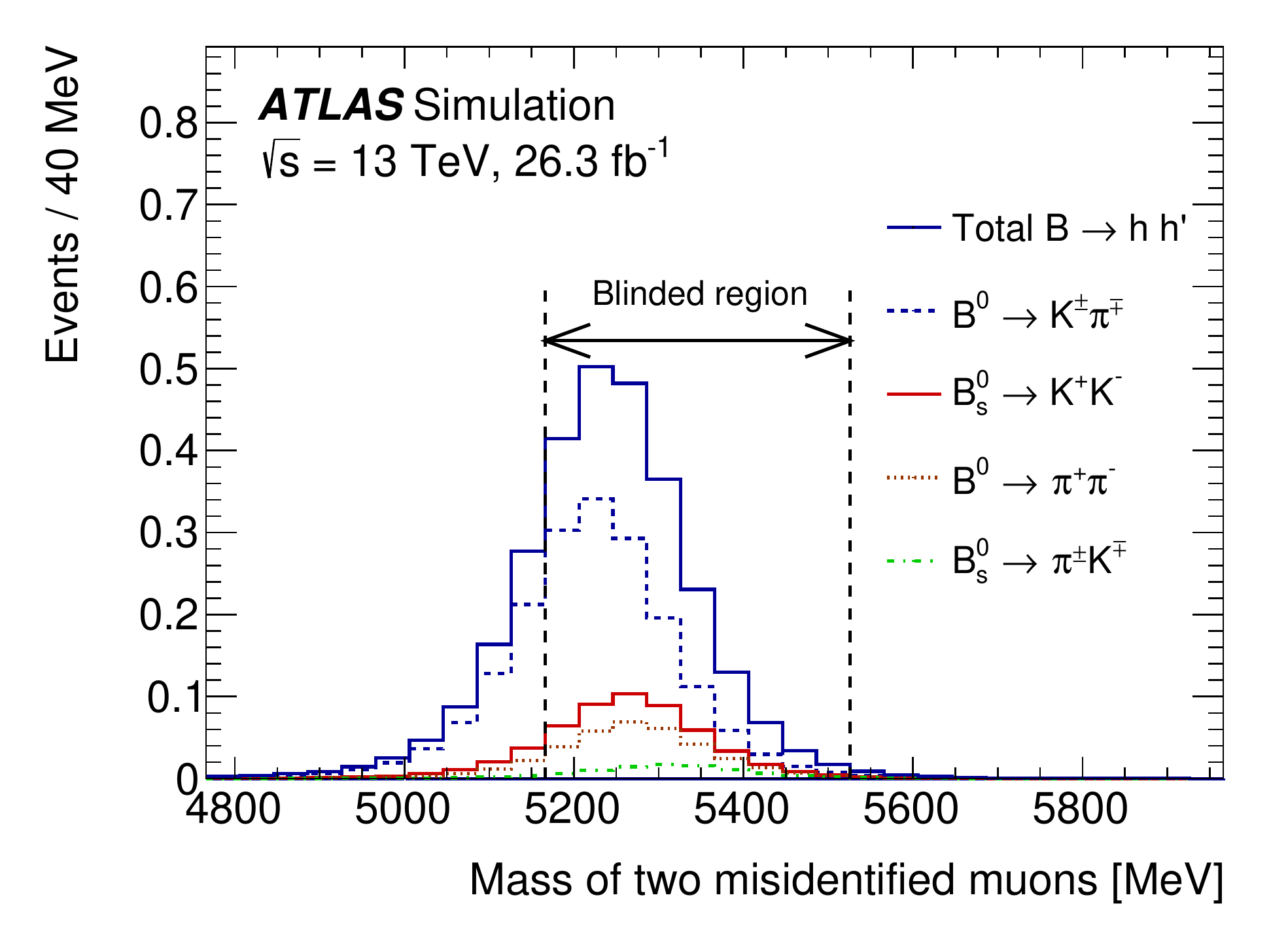}
\hspace*{-0.6cm}
}
\caption{(a) Dimuon invariant mass distribution for the partially reconstructed
background (as categorised in Section~\ref{sec:bkg}), from simulation, before the final selection against continuum is
applied but after all other requirements.
The different components are shown as stacked histograms, normalised according
to world-averaged measured branching fractions. The SM expectations for the
\Bmumu\ signals are also shown for comparison. Continuum background is not included here.
(b) Invariant mass distribution of the \Bhh peaking background components
after the complete signal selection is applied. The $B^0_s \to \pi^+\pi^-$ and
$B^0 \to K^+ K^-$ contributions are negligible on this scale.
In both plots the vertical dashed lines indicate the blinded analysis region. Distributions are normalised to the expected yield for the
integrated luminosity of 26.3 fb$^{-1}$.}
\label{fig:SS-SV}
\end{center}
\end{figure}
The semileptonic decays with final-state hadrons misidentified as muons
consist mainly of three-body charmless decays $B^0 \to \pi^-\mu^+\nu$,
$B^0_s \to K^-\mu^+\nu$ and $\Lambda_b \to p\mu^-\overline{\nu}$ in which the tail
of the invariant mass distribution extends into the signal region.
Due to branching fractions of the order of $10^{-6}$,
this background is not large, and is further reduced by the muon identification
requirements, discussed in Section~\ref{sec:fakes}.
The MC invariant mass distributions of these partially reconstructed decay topologies are shown
together with the SM signal predictions in Figure~\ref{fig:SS-SV}(a) after applying the preliminary selection criteria
described in Section~\ref{sec:sel}.
 
Finally, the peaking background is due to \Bhh\ decays containing
two hadrons misidentified as muons.  The distributions in Figure~\ref{fig:SS-SV}(b),
obtained from simulation, show that these decays populate the signal region.
This component is further discussed in Section~\ref{sec:fakes}.
\section{Hadron misidentification}
\label{sec:fakes}
 
In the preliminary selection, muon candidates are formed from the combination
of tracks reconstructed independently in the ID and MS.
The performance of the muon reconstruction in ATLAS is presented in Ref.~\cite{PERF-2015-10}.
Additional studies were performed to
evaluate the amount of background related to hadrons erroneously identified as muons.
 
Detailed simulation studies were performed for the \Bhh\ channel with a
full \Geant-based simulation~\cite{GEANT4} of all systems of the ATLAS detector.
The vast majority of background events from particle misidentification are due to decays
in flight of kaons and pions, in which the muon receives most of the energy of the parent meson.
Hence this background is generally related
to true muons measured in the MS, but not produced promptly in the decay of a $B$ meson.

The muon candidate is required to pass {\em tight} muon requirements in the preliminary selection,
which are based on the profile of energy deposits in the calorimeters as well as
on tighter ID--MS matching criteria than those used for the \textit{loose} requirements.
Two-body $B$ decays in control regions show that {\em tight} selections have, relative to the {\em loose} counterpart, an
average hadron misidentification probability reduced by a factor 0.39 with a muon reconstruction efficiency of 90\%.
The resulting final value of the misidentification probability is
0.08\% for kaons and 0.1\% for pions.
Efficiencies and fake rates are relative to the analysis
preselections, including tracking but excluding any muon requirement.
 
The background due to \Bhh, with double misidentification of $h h^\prime$ as $\mumu$,
has a reconstructed invariant mass distribution that peaks at 5240~\MeV, close to the
\Bz\ mass, and is effectively indistinguishable from the \Bz\ signal
(see Figure~\ref{fig:SS-SV}(b)).
The expected number of peaking-background events can be estimated in a way analogous to that for the signal, from the number of observed
\BpKpJpsi\ events using Eq.~(\ref{eq:BRFormula}), after taking into account the expected differences from muon identification variables
and trigger selections.
World average~\cite{PDG2018} values for the branching fractions of \Bz\ and \Bs\ into
$K\pi$, $KK$ and $\pi\pi$ are used, together with the hadron
misidentification probabilities obtained from simulation.  This results in $2.7\pm 1.3$ total
expected peaking-background events, after the reference multivariate selection.\footnote{This selection, corresponding to 54\% signal efficiency,
was also applied to derive all other quantities quoted in this Section and includes a selection against the $\mumu$ continuum background
based on the BDT discussed in Section~\ref{sec:contBDT}.}
 
When selecting {\em loose} muons and inverting the additional requirements imposed in the
{\em tight} muon selection, the number of events containing real muons is substantially reduced,
while the number of peaking-background events is approximately two times
larger than in the sample obtained with the nominal selection.
A fit to data for this background-enhanced sample
returns $6.8 \pm 3.7$ events, which translates into a
peaking-background yield in the signal region of $2.9 \pm 2.0$ events
when taking into account the relative rejection of the muon
quality selections. The predicted yield is in good agreement with the simulation.
 
Besides the peaking background, the {\em tight} muon selection also reduces the
semileptonic contributions with a single misidentified hadron. Simulation yields $30 \pm 3$ events expected
from $\Bz \to \pi^-\mu^+\nu$ and $\Bs \to K^-\mu^+\nu$ in the final
sample, with a distribution kinematically constrained to be mostly below the signal region.
The $\Lambda_b \to p\mu^-\overline{\nu}$ contribution is negligible due to the smaller production cross section
and the low rate at which protons fake muons.

\section{Continuum background reduction}
\label{sec:contBDT}
 
A multivariate analysis, implemented as a boosted decision tree (BDT), is employed
to enhance the signal relative to the continuum background.
This BDT is based on the $15$ variables described in Table~\ref{tab:15vars}.
\begin{table}[htb]
\small
\begin{center}
\caption{Description of the $15$ input variables used in a BDT classifier to discriminate between signal
and continuum background. When the BDT classifier is applied to \BpKpJpsi\ and
\BsJpsiPhi\ candidates, the variables related to the decay products of the $B$
mesons refer only to the muons from the decay of the $J/\psi$. Horizontal lines
separate the classifications into groups (a), (b) and (c) respectively, as described in the text.
For category (c), additional tracks are required to have \pt>500 \MeV.}
\begin{tabular}{p{2cm} p{13cm}}
\toprule \\[-9pt]
Variable  &  Description \\[2pt]
\midrule \\[-9pt]
$\pt^{B}$ & Magnitude of the $B$ candidate transverse momentum $\overrightarrow{\pt}^B$.\\[5pt]
$\chi^{2}_{{\mathrm{PV,DV}}\;xy}$ & Compatibility of the separation
$\overrightarrow{\Delta x}$ between production (i.e. associated PV) and decay (DV)
vertices in the transverse projection:
$\overrightarrow{\Delta x}_\text{T}\!\cdot \! \Sigma_{\overrightarrow{\Delta x}_\text{T}}^{\phantom{x}-1}\!\cdot\!\overrightarrow{\Delta x}_\text{T}$,
where $\Sigma_{\overrightarrow{\Delta x}_\text{T}}$ is the covariance matrix.
\\[5pt]
$\Delta{R}_{\mathrm{flight}}$ & Three-dimensional angular distance between $\overrightarrow{p}^{B}$ and
$\overrightarrow{\Delta x}$: $\sqrt{\alpha_{2\mathrm{D}}{}^{2} + (\Delta\eta)^{2}}$
\\[3pt]
$|\alpha_{2\mathrm{D}}|$ & Absolute value of the angle in the transverse plane between $\overrightarrow{\pt}^{B}$ and
$\overrightarrow{\Delta x}_\text{T}$.
\\[3pt]
$L_{xy}$ & Projection of $\overrightarrow{\Delta x}_\text{T}$ along the direction of $\overrightarrow{p}_\text{T}^{B}$:
$(\overrightarrow{\Delta x}_\text{T}\!\cdot\!\overrightarrow{\pt}^{B})/|\overrightarrow{\pt}^{B}|$. \\[3pt]
IP$_B^{3\mathrm{D}}$ & Three-dimensional impact parameter
of the $B$ candidate to the associated PV.\\[3pt]
\midrule \\[-9pt]
DOCA$_{\mu\mu}$ & Distance of closest approach (DOCA) of the two tracks forming the $B$ candidate (three-di\-men\-sion\-al).
\\[3pt]
$\Delta\phi_{\mu\mu}$ & Azimuthal angle between the momenta of the two tracks forming the $B$ candidate.
\\[3pt]
$|d_{0}|^{\mathrm{max}}$-sig. & Significance of the larger absolute value of the impact
parameters to the PV of the tracks forming the $B$ candidate,  in the transverse plane.
\\[3pt]
$|d_{0}|^{\mathrm{min}}$-sig. & Significance of the smaller absolute value of the impact
parameters to the PV of the tracks forming the $B$ candidate, in the transverse plane.
\\[3pt]
$P_{\text{L}}^{\mathrm{min}}$ & The smaller of the projected values of the muon momenta
along $\overrightarrow{\pt}^B$.
\\[3pt]
\midrule \\[-9pt]
$I_{0.7}$
& Isolation variable defined as ratio of $|\overrightarrow{\pt}^{B}|$ to the sum of $|\overrightarrow{\pt}^{B}|$ and
the transverse momenta of all additional tracks contained
within a cone of size $\Delta{R}=\sqrt{(\Delta\phi)^2+(\Delta\eta)^2} = 0.7$ around the $B$ direction.
Only tracks matched to the same PV as the $B$ candidate are included in the sum.
\\[3pt]
DOCA$_{\mathrm{xtrk}}$ & DOCA of the closest additional track to the decay vertex of the $B$ candidate.
Only tracks matched to the same PV as the $B$ candidate are considered.\\[3pt]
$N^{\mathrm{close}}_{\mathrm{xtrk}}$ & Number of additional tracks compatible with the decay vertex (DV) of the
$B$ candidate  with $\ln(\chi^2_{\mathrm{xtrk,DV}})\! < \! 1$.
Only tracks matched to the same PV as the $B$ candidate are considered.
\\[3pt]
$\chi^2_{\mu,\mathrm{xPV}}$ & Minimum $\chi^2$ for the compatibility of a muon in the $B$ candidate
with any PV reconstructed in the event.\\[5pt]
\bottomrule
\end{tabular}
\label{tab:15vars}
\end{center}
\end{table}
The discriminating variables can be classified into three groups:
(a) $B$ meson variables, related to the reconstruction of the decay vertex and
to the collinearity between $\overrightarrow{p}^B$ and the flight vector between
the production and decay vertices $\overrightarrow{\Delta x}$;
(b) variables describing the muons that form the $B$ meson candidate;
and (c) variables related to the rest of the event.
The selection of the variables aims to maximise the discrimination power
of the classifier without introducing significant dependence on the invariant mass
of the muon pair.
 
The same discriminating variables were used in the previous
analysis based on the full Run~1 dataset~\cite{Aaboud:2016ire}.
The removal of individual variables was explored to simplify the BDT input,
however, this results inevitably in a significant reduction of the BDT separation power.
To minimise the dependence of the classifier on the effects of pile-up,
the additional tracks considered to compute the variables
$I_{0.7}$, DOCA$_{\mathrm{xtrk}}$ and $N^{\mathrm{close}}_{\mathrm{xtrk}}$
are required to be compatible with the primary vertex matched to the dimuon candidate.
 
The correlations among the discriminating variables
were studied in the MC samples for signal and continuum background
discussed in Section~\ref{sec:data},
and in data from the sidebands of the $\mumu$ invariant mass distribution.
There are significant linear correlations among the variables
$\chi^{2}_{{\mathrm{PV,DV}}\,xy}$, $L_{xy}$, $|d_{0}|^{\mathrm{max}}$-sig.,
$|d_{0}|^{\mathrm{min}}$-sig. and $\chi^2_{\mu,\mathrm{xPV}}$.
The variables IP$_B^{3\mathrm{D}}$, DOCA$_{\mu\mu}$ and $I_{0.7}$ have negligible correlation
with any of the others used in the classifier.
 
The simulated signal sample and the data from the dimuon invariant mass sideband regions are used
for training and testing the classifier.
As discussed in Section~\ref{sec:data}, simulated signal samples are corrected for muon
reconstruction efficiency differencies between simulation and data, and reweighted according
to the distributions of \pt and $|\eta|$ of the dimuon and of the pile-up observed in data.
The BDT training is done using the TMVA toolkit~\cite{TmvaManual}.
 
Sideband data are used for the BDT training and optimisation. The sample is subdivided into
three randomly selected separate and equally populated subsamples used
in turn to train and validate the selection efficiency of three independent BDTs.
The resulting BDTs are found to produce results that are statistically compatible, and are combined in one
single classifier in such a way that each BDT is applied only to the part of the data sample not involved in the
BDT training.
 
\begin{figure}[!tb]
\begin{center}
\subfigure[]{
\hspace*{-0.5cm}
\includegraphics[width=0.49\textwidth]{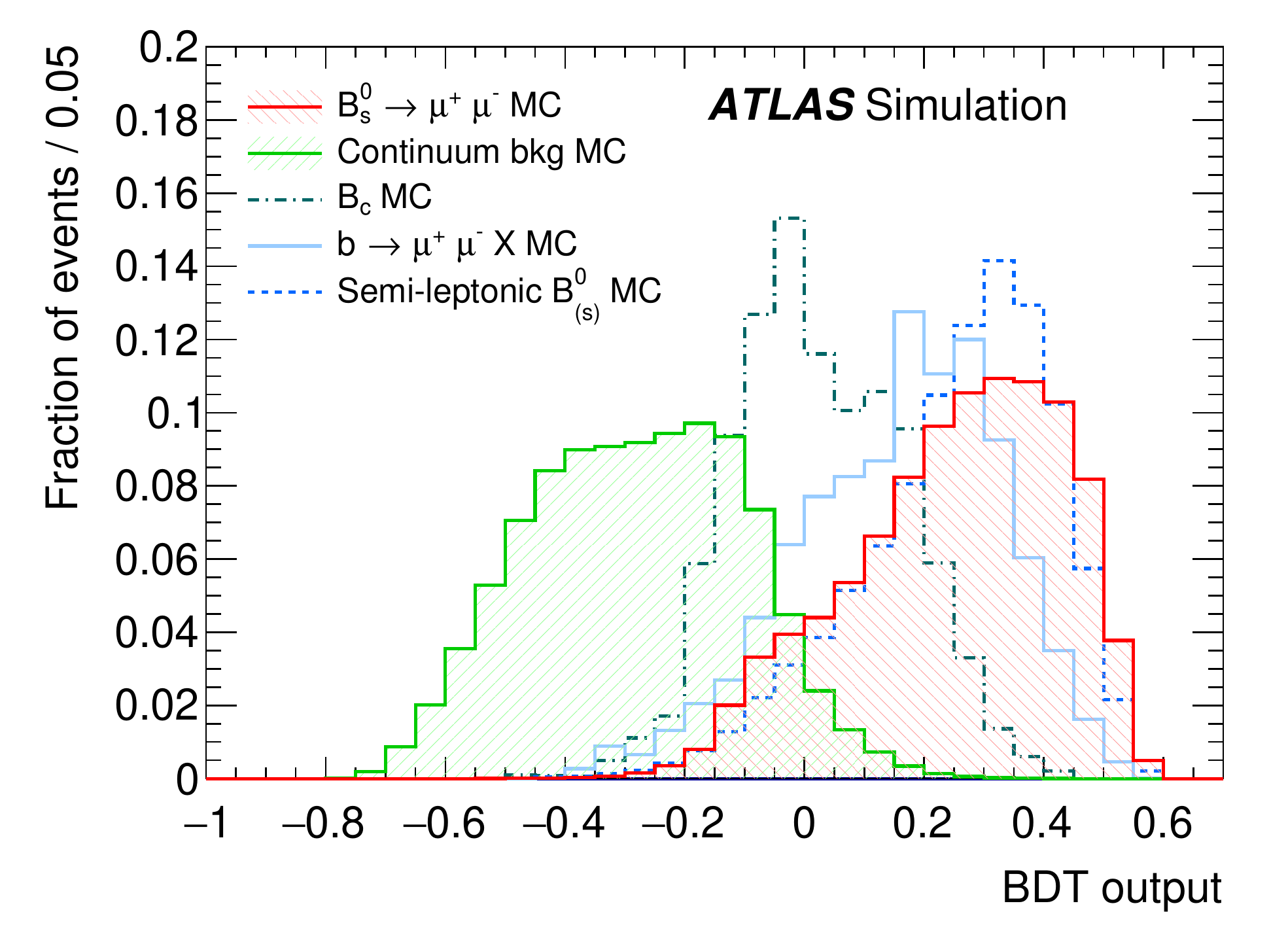}
}
\subfigure[]{
\hspace*{-0.5cm}
\includegraphics[width=0.49\textwidth]{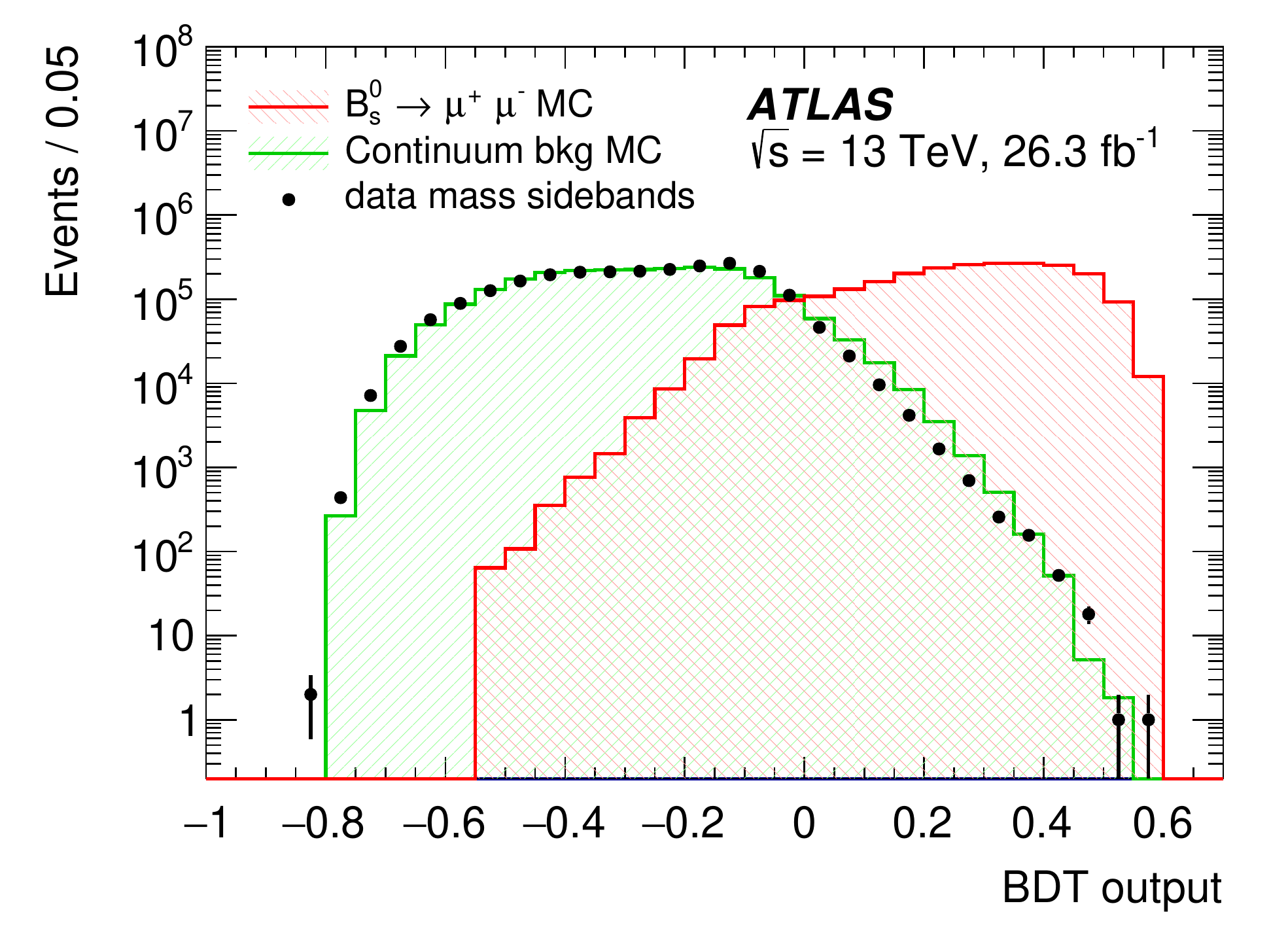}
}
\caption{
BDT output distribution for the signal and background events after the preliminary selection and before applying any reweighting to the BDT input variables:
(a) simulation distributions for \Bsmumu signal, continuum, partially reconstructed $b\to \mumu X$ events and
\Bc\ decays; (b) dimuon sideband candidates (which also include prompt contributions,
mainly at lower BDT values and not simulated in the continuum MC sample), compared with the continuum MC sample and the simulated signal.
All distributions are normalised to unity in (a) and to data sidebands in (b).
}
\label{fig:contBDT}
\end{center}
\end{figure}

Figure~\ref{fig:contBDT} shows the distribution of the BDT output variable
for simulated signal and backgrounds, separately for continuum background and partially reconstructed events.
Also shown is the BDT distribution for dimuon candidates from the sidebands of the
invariant mass distribution in data.
The BDT output was found to not have any significant correlation with the dimuon invariant mass.
The final selection requires a BDT output value larger than $0.1439$, corresponding to signal
and continuum background efficiencies of 72\% and 0.3\% respectively. The analysis
uses all candidates after this selection; however, accepted events with BDT values close to the selection threshold
are effectively only constraining the background models.\footnote{The \Bmumu signal fit was found to be
insensitive to the signal for candidates with BDT output value smaller than 0.2455 (corresponding
to 54\% and 0.03\% efficiencies for signal and background respectively).}
For this reason, signal and reference channel yields and efficiencies are measured
relative to the \textit{signal reference} selection discussed in Section~\ref{sec:eff}, while the events in the final selection with lower BDT values are used to improve the background modelling.

\section{Data--simulation comparisons}
\label{sec:comp}

\begin{figure}[!htb]
\begin{center}
\subfigure[]{
\hspace*{-0.5cm}
\includegraphics[width=0.52\textwidth]{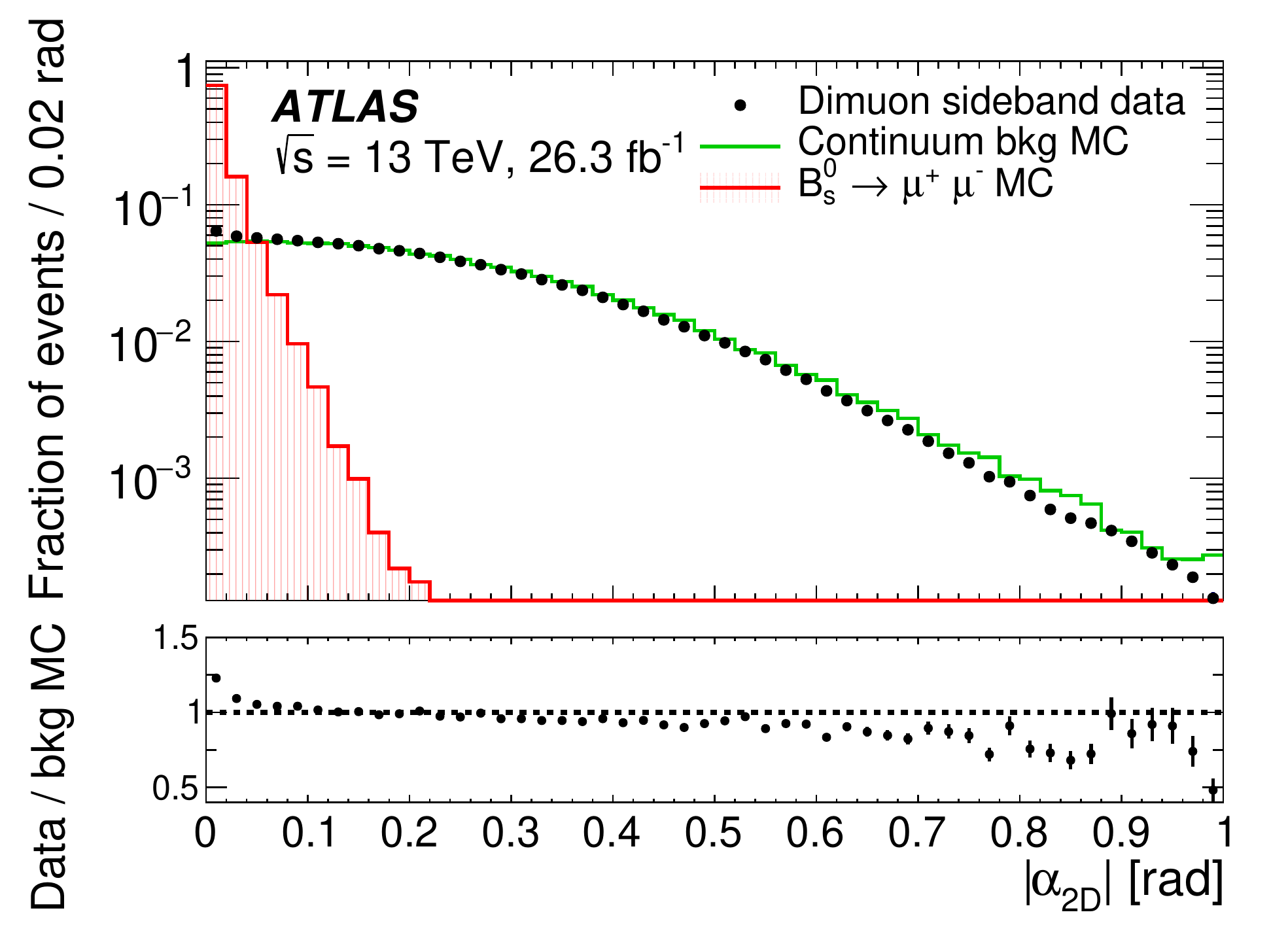}
}
\subfigure[]{
\hspace*{-0.5cm}
\includegraphics[width=0.52\textwidth]{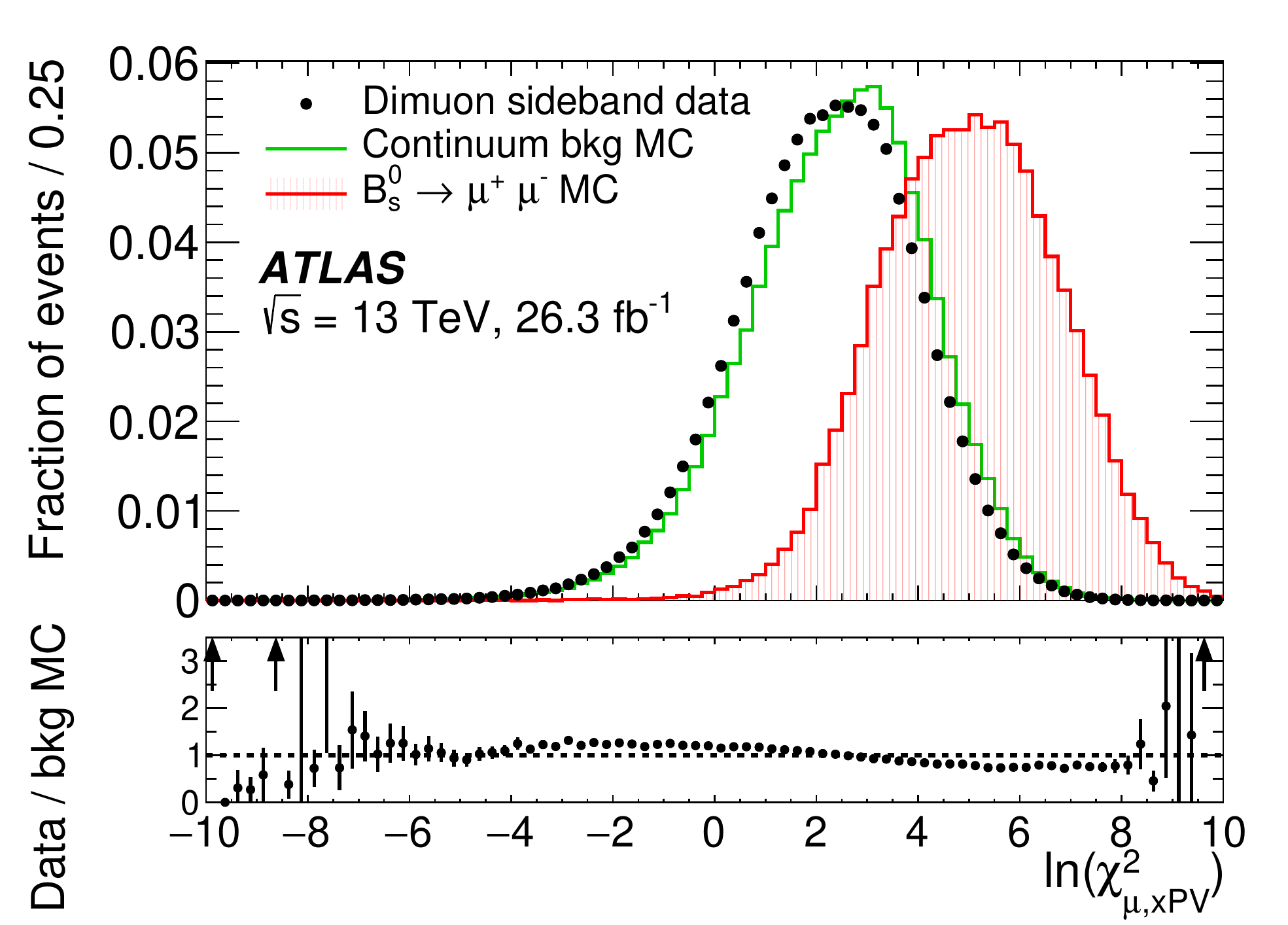}
\hspace*{-0.6cm}
}
\caption{Data and continuum MC distributions of the (a) $|\alpha_{2\mathrm{D}}|$
and (b) $\ln \left(\chi^2{}_{\mu,\mathrm{xPV}}\right)$ variables (defined in Table~\ref{tab:15vars}).
The points correspond to the sideband data, while
the continuous-line histogram corresponds to
the continuum MC distribution, normalised to the number of data events.
The filled-area histogram shows the signal MC distribution
for comparison. The bottom insets report the data/MC ratio, zoomed-in in order
to highlight discrepancies in the region that is most relevant for the analysis.}
\label{fig:maincompcont}
\end{center}
\end{figure}

Despite the lack of correlation between the BDT variable and the candidates invariant mass, the change in relative contribution of different background components as a function of the BDT output can produce a variation of the background shape in separate BDT bins.
The shape parameters and normalisations of the backgrounds are for this reason determined purely from data: the $b\bar{b}\to\mumu X$ MC simulation is not used for BDT training, computation of efficiencies or normalisation purposes.
The only role of the $b\bar{b}\to\mumu X$ MC is the validation of the functional forms employed to parameterise the backgrounds on data, as discussed in Section \ref{sub:sec:model}. The dependence of the background parameterisation on the BDT output variable was investigated in data and shown to be consistent with simulation, even for lower BDT values where prompt backgrounds contribute significantly.
This observation is supported by the fair agreement of the BDT input variables distribution between the simulated $b\bar{b}\to\mumu X$ background and what observed on data sidebands.
Figure \ref{fig:maincompcont} compares the distributions of two of the most discriminating variables in the continuum background MC sample with data in the dimuon mass sidebands. Agreement with the sideband data is fair.
 
\begin{figure}[!h]
\begin{center}
\subfigure[]{
\hspace*{-0.5cm}
\includegraphics[width=0.52\textwidth]{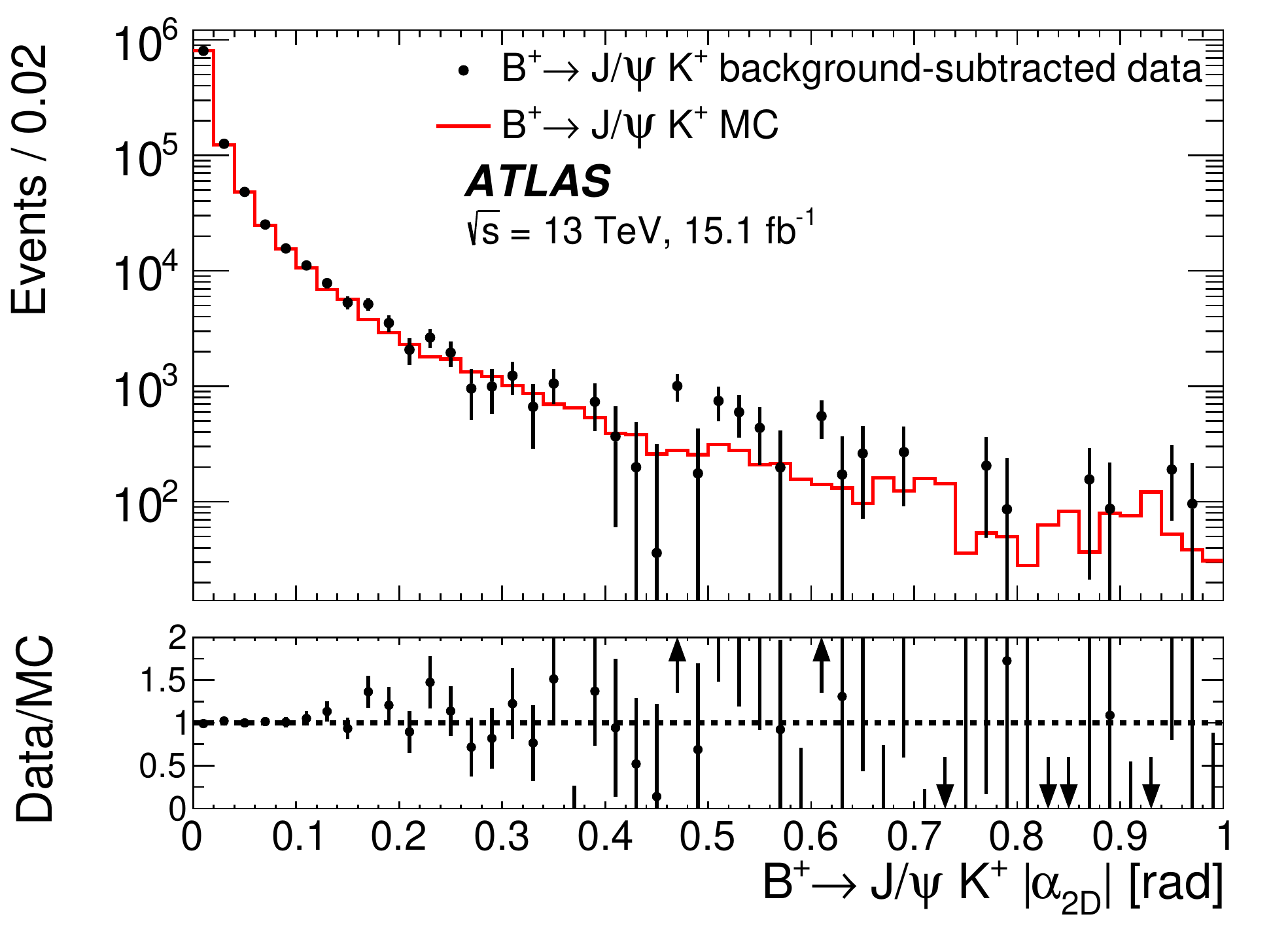}
}
\subfigure[]{
\hspace*{-0.5cm}
\includegraphics[width=0.52\textwidth]{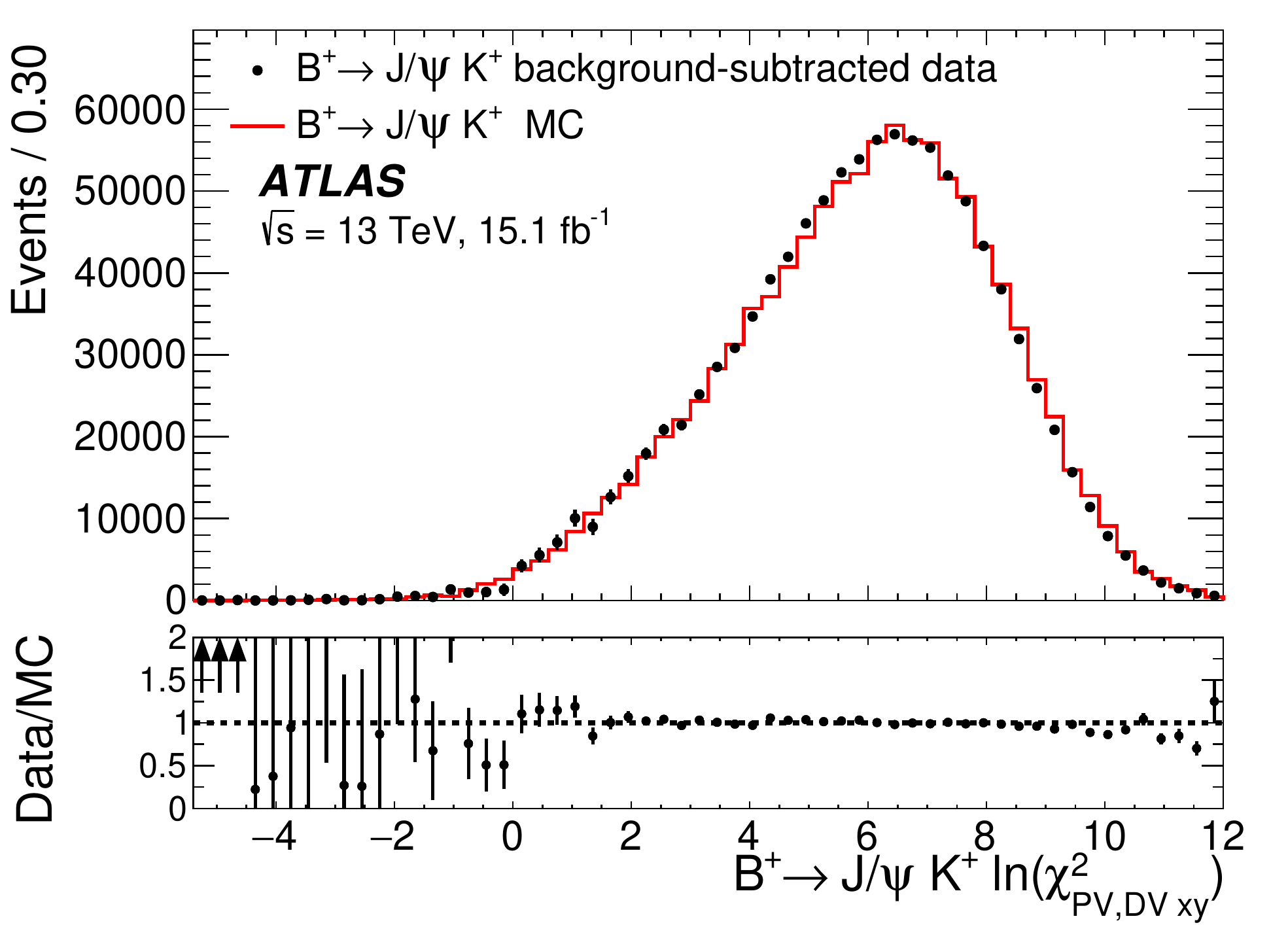}
\hspace*{-0.6cm}
}\\
\subfigure[]{
\hspace*{-0.5cm}
\includegraphics[width=0.52\textwidth]{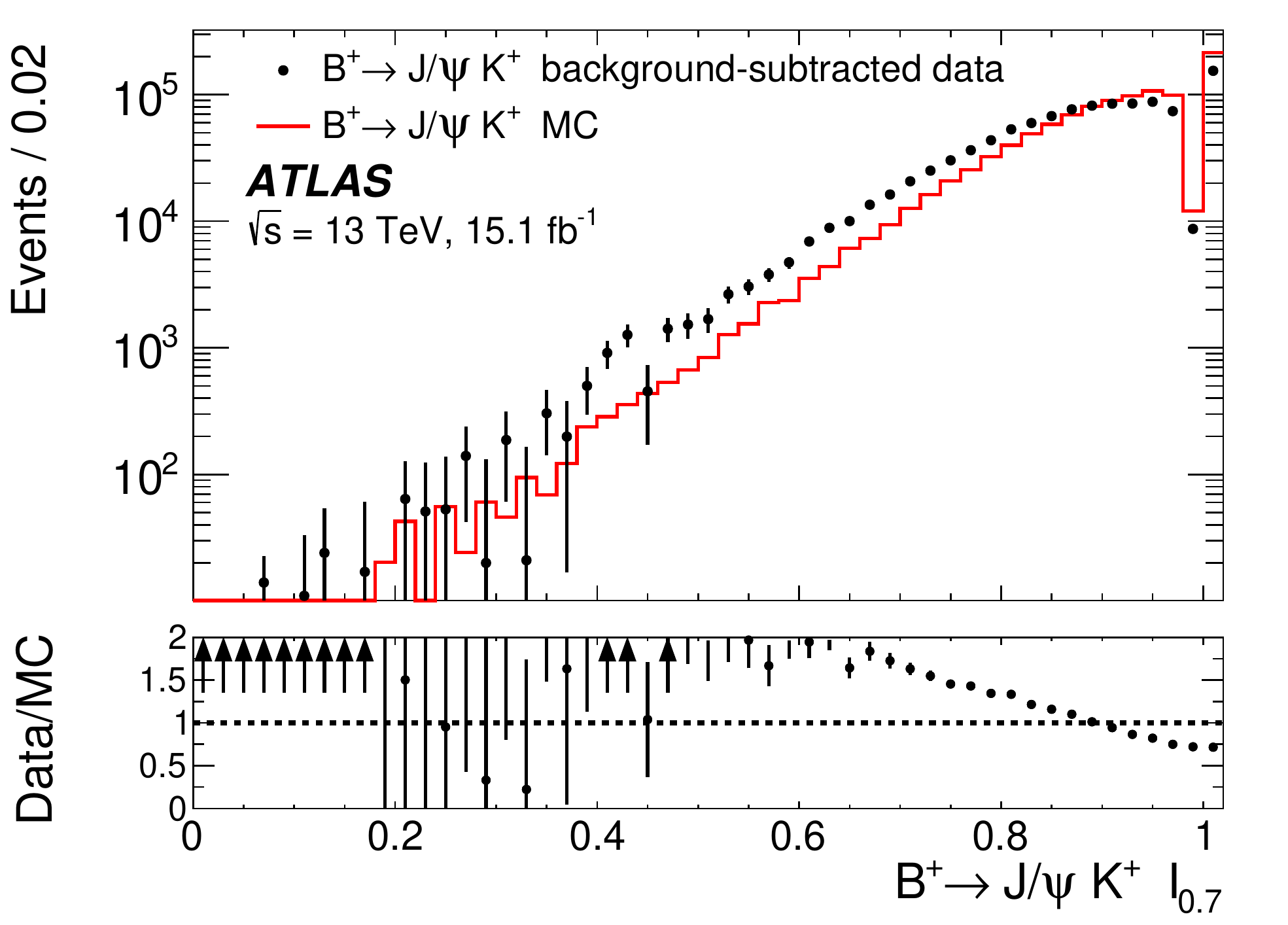}
}
\subfigure[]{
\hspace*{-0.5cm}
\includegraphics[width=0.52\textwidth]{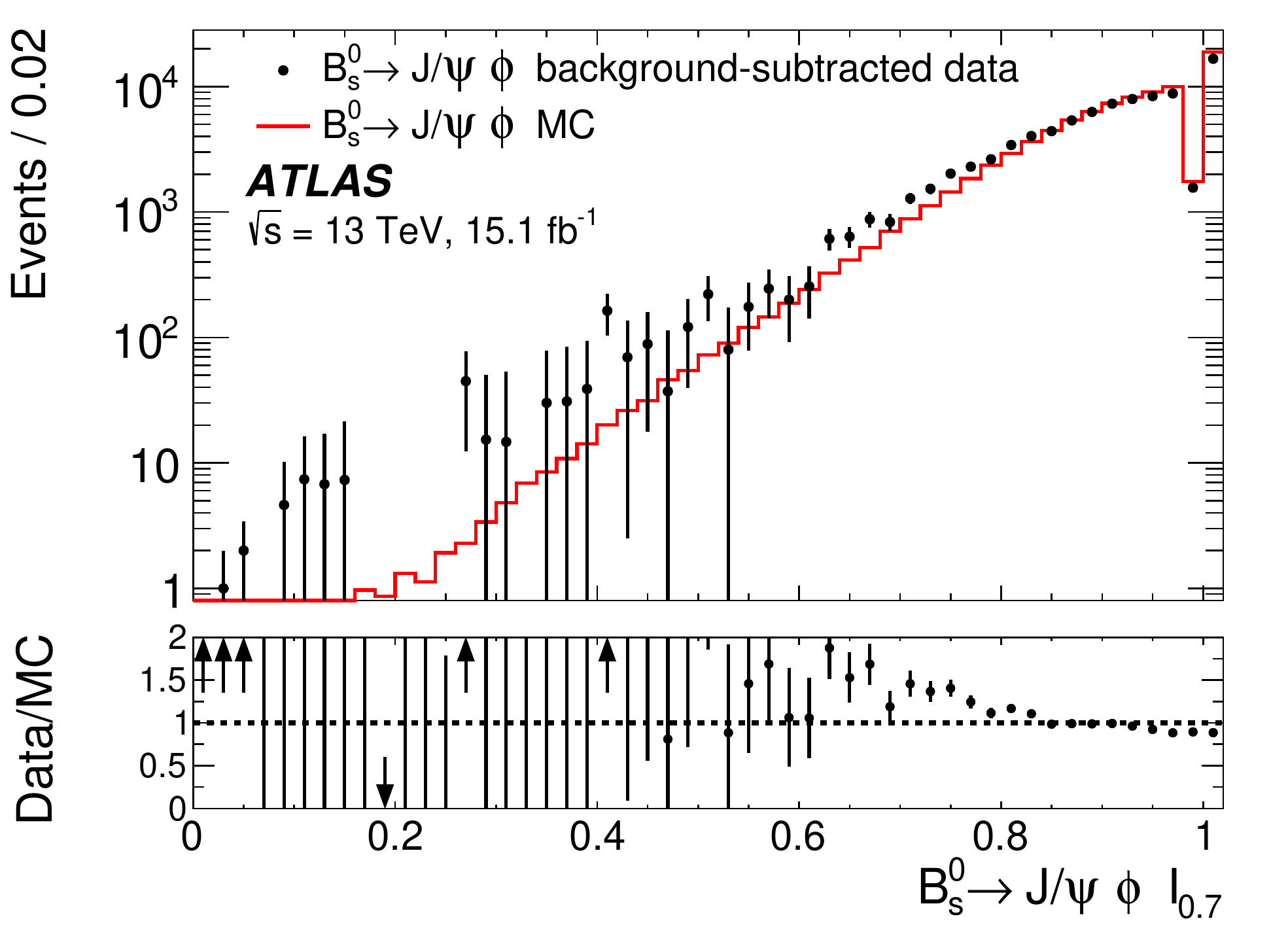}
\hspace*{-0.6cm}
}
\caption{
Data and MC distributions in \BpKpJpsi\ events for
the discriminating variables: (a) $|\alpha_{2\mathrm{D}}|$,
(b) $\ln\left(\chi^{2}_{{\mathrm{PV,DV}}\;xy}\right)$ and (c) $I_{0.7}$.
The variable $I_{0.7}$ is also shown in (d) for \BsJpsiPhi\ events.
The points correspond to the sideband-subtracted data, while
the line corresponds to
the MC distribution, normalised to the number of data events.
The highest bin in (c) and (d) accounts for the events with $I_{0.7}=1$.
The bottom insets report the data/MC ratio, zoomed-in in order to highlight
discrepancies in the region that is most relevant for the analysis.
}
\label{fig:discri-var-comparison}
\end{center}
\end{figure}
 
The distributions of the discriminating variables are also used to compare simulation
and data in the \BpKpJpsi\ and \BsJpsiPhi\ samples.
To perform these comparisons,
for each variable the contribution of the background is subtracted from the
\BpKpJpsi\ (\BsJpsiPhi) signal.
For this purpose, a maximum-likelihood fit is performed to the invariant mass distribution,
separately in bins of rapidity and transverse momentum.
The fit model used is simpler than the one employed for the extraction of the \Bp\ signal
for normalisation as described in Section~\ref{sec:bplus},
but is sufficient for the purpose discussed here.
 
Figure~\ref{fig:discri-var-comparison} shows examples of the distributions of
the discriminating variables obtained from data and simulation for the reference samples.
Observed differences are used to estimate systematic uncertainties on
the efficiency ratio $\theRho=\varepsilon(\BpKpJpsi)/\varepsilon(\Bmumu)$
with the procedure described in Section~\ref{sec:eff}.
The discrepancy visible for the isolation variable $I_{0.7}$ in the \BpKpJpsi\ channel
is the most significant among all variables and both reference channels.
 
\section{$B^+ \rightarrow J/\psi K^+$ yield extraction}
\label{sec:bplus}
 
The reference channel yield is extracted with an
unbinned extended maximum-likelihood fit to the $J/\psi K^+$ invariant mass distribution.
The functional forms used to model both the signal and the backgrounds are obtained from studies of MC samples.
All the yields are extracted from the fit to data, while the shape parameters are determined from a simultaneous fit to data and MC samples.
Free parameters are introduced for the mass scale and
mass resolution to accommodate data--MC differences. The best-fit values indicate
a negligibly poorer resolution and a mass shift at the level of $2$~\MeV.
 
\begin{figure}[!htb]
\begin{center}
\includegraphics[width=0.6\textwidth]{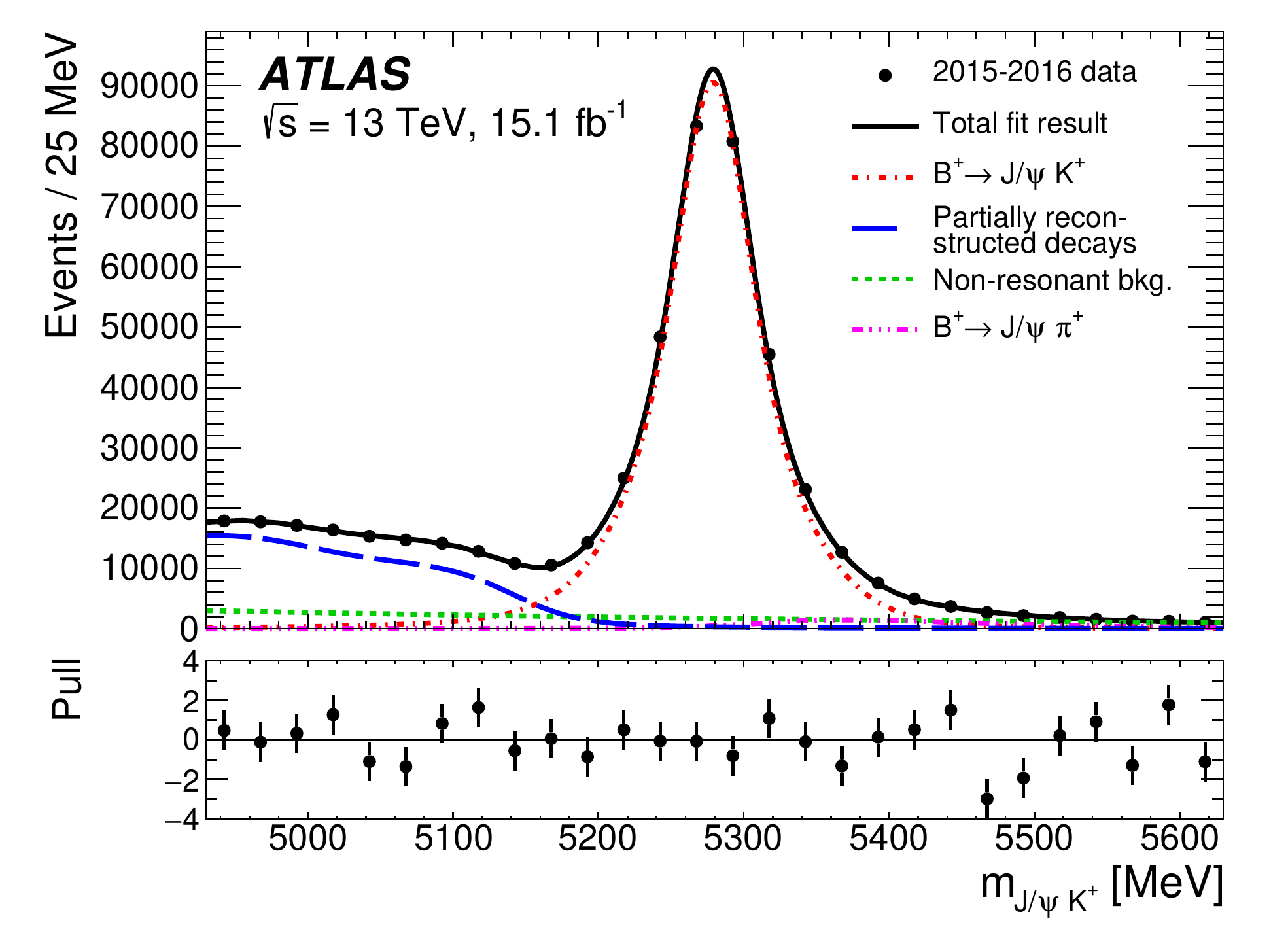}
\caption{Result of the fit to the $J/\psi K^+$ invariant mass distribution for all $B^+$
candidates in half of the data events. 
The various components of the spectrum are described in the text.
The inset at the bottom of the plot shows the bin-by-bin pulls for the
fit, where the pull is defined as the difference between the data point
and the value obtained from the fit function, divided by the error from the fit.
}
\label{fig:bplus}
\end{center}
\end{figure}
 
The fit includes four components: \BpKpJpsi\ decays, Cabibbo-suppressed
\BpPipJpsi\ decays in the right tail of the main peak, partially reconstructed
$B$ decays (PRD) where one or more of the final-state particles are missing,
and the non-resonant background composed mostly of $b\bar{b}\to \jpsi X$ decays.
All components other than the last one have shapes constrained by MC simulation as described below,
with the data fit including an additional Gaussian convolution to account for possible data--MC
discrepancies in mass scale and resolution.
The shape of the \BpKpJpsi\ mass distribution is parameterised using a Johnson $S_U$
function~\cite{Johnson-1,Johnson}.
The final \BpKpJpsi\ yield includes the contribution from radiative 
effects (i.e. where photons are emitted from the $B$ decay products).
The \BpPipJpsi\ decays are modelled by the sum of a Johnson $S_U$ function and a Gaussian
function, where all parameters except the normalisation are determined from the simulation.
The decay modes contributing to the PRD are classified in simulation on the basis of their mass dependence.
Each of the three resulting categories contributes to the overall PRD shape with combinations of Fermi--Dirac and exponential functions,
contributing differently in the low-mass region.
Their shape parameters are determined from simulation.
Finally, the non-resonant background is modelled with an exponential function
with the shape parameter extracted from the fit.
The normalisation of each component is unconstrained in the fit, which is therefore
mostly independent of external inputs for the branching fractions. The residual dependence
of the PRD model shapes on the relative branching fractions of the contributing decays
is considered as a source of systematic uncertainty.
The resulting fit, shown in Figure~\ref{fig:bplus}, yields $334\,351$ \BpKpJpsi\ decays with
a statistical uncertainty of 0.3\%. The ratio of yields of \BpPipJpsi\ and \BpKpJpsi\ is
$\left(3.71 \pm 0.09\right)$\% (where the uncertainty reported is statistical only), in agreement with the expectation from the world
average~\cite{PDG2018} of $\left(3.84 \pm 0.16\right)$\%.
 
Some systematic uncertainties are included 
by design in the fit.
For example, the effect of the limited MC sample size
is included
by performing a simultaneous fit to data and MC samples.
Scaling factors determined in the fit to data account for the differences in mass scale
and resolution between data and simulation.
Additional systematic uncertainties are evaluated by varying the default
fit model described above.  They take into account the kinematic differences
between data and the MC samples used in the fit,
differences in efficiency between $B^+$ and $B^-$ decays and
uncertainties in the relative fractions and shapes of PRD and in the shape
of the various fit components.
The stability of this large sample fit
is verified by repeating the fit with different initial parameter values.
In each case, the change relative to the default fit is recorded,
symmetrised and used as an estimate of the systematic uncertainty.
The main contributions to the systematic uncertainty come from the functional models
of the background components, the
composition
of the PRD and the signal charge asymmetry.
The total systematic uncertainty in the $B^+$ yield amounts to $4.8\%$.
 
\section{Evaluation of the \BpKpJpsi\ to \Bmumu\ efficiency ratio}
\label{sec:eff}
The ratio of efficiencies $\theRho=\varepsilon(\BpKpJpsi)/\varepsilon(\Bmumu)$ enters the
$\mathcal{D}_{\mathrm{ref}}$ term defined in Section~\ref{sec:outline}: $\mathcal{D}_{\mathrm{ref}}=N_{J/\psi K^+}/\theRho$.
Both channels are measured in the fiducial acceptance for
the $B$ meson,
defined as $\pt^B>8.0$~\GeV\ and $\left| \eta_B\right|<2.5$.
Correspondingly, $\varepsilon(\BpKpJpsi)$ and $\varepsilon(\Bmumu)$ are measured within the $B$ meson
fiducial acceptance and include additional final state particles
acceptance as well as trigger, reconstruction and selection efficiencies.
The final state particles acceptance is defined by the selection placed on the particles in the final state:
$|\eta_{\mu}|<2.5$ and $\pt^{\mu}>6.0$ $(4.0)$~\GeV\ for the leading (trailing)~muon \pt,
$\pt^{K}>1.0$~\GeV\ and $\left| \eta_{K}\right|<2.5$ for kaons.
The \textit{signal reference} BDT selection, defined as BDT~$>$~0.2455,
has an efficiency of about $54\%$ ($51\%$) in the signal (reference) channel.
The overall efficiency ratio \theRho is 0.1176$\,\pm\,$0.0009~(stat.)$\,\pm\,$0.0047~(syst.), with
uncertainties determined as described below.
 
The ratio \theRho is computed using the mean lifetime of $\Bs$~\cite{HFLAV2016, PDG2018}
in the MC generator. The same efficiency ratios apply to the \Bsmumu\ and \Bdmumu\
decays, within the MC statistical uncertainty of $0.8\%$.
The statistical uncertainties in the efficiency ratios come from the finite number of
events available for the simulated samples.
The systematic uncertainty affecting \theRho\ comes from five sources.
 
The first contribution is due to the uncertainties in the data-driven weights introduced in Section~\ref{sec:data},
and amounts to 0.8\%. This term is assessed
by creating alternative datasets using
correction factors that are randomly sampled in accord with their nominal values and
uncertainties. The RMS value of the distribution of \theRho\ obtained from
these datasets
is taken as the systematic uncertainty.
 
A second contribution of 1.0\% is related to the muon trigger and reconstruction efficiencies.
The effect of the uncertainties in the data-driven efficiencies is evaluated
using random sampling, as above.
 
A 3.2\% systematic uncertainty contribution arises from the differences
between data and simulation observed in the modelling of the discriminating variables
used in the BDT classifier (Table~\ref{tab:15vars}). For each of the $15$ variables, the
MC samples for \Bmumu\ and \BpKpJpsi\ are reweighted with the ratio of the \BpKpJpsi\ event distributions in
sideband-subtracted data and the MC simulation.
The isolation variable $I_{0.7}$ is computed using charged-particle tracks only, and differences
between \Bp\ and \Bs\ are expected and were observed in previous studies~\cite{bsmumuv1}.
Hence for this variable the reweighting procedure for the \Bsmumu\ MC sample is based
on \BsJpsiPhi\ data.
For all discriminating variables except $I_{0.7}$,
the value of the efficiency ratio is modified by less than 2\% by the reweighting procedure
and each variation is taken as an independent contribution to the systematic
uncertainty in the efficiency ratio. For $I_{0.7}$ the reweighting procedure changes
the efficiency ratio by about $6\%$. Because of the significant mis-modelling, the MC samples
obtained after reweighting on the distribution of $I_{0.7}$ are taken as a reference,
thus correcting the central value of the efficiency ratio.
The 1\% uncertainty in the $I_{0.7}$ correction is added to the sum in quadrature
of the uncertainties assigned to the other discriminating variables.
The total uncertainty in the modelling of the discriminating variables is the dominant
contribution to the systematic uncertainty in \theRho.
 
A fourth source of systematic uncertainty arises
from differences between the \Bsmumu\ and the \BpKpJpsi\ channel
related to the reconstruction efficiency of the kaon track
and of the \Bp\ decay vertex.
These uncertainties are mainly due to inaccuracy in the modelling
of passive material in the ID. The corresponding systematic uncertainty is
estimated by varying the detector model in simulations,
which results in changes between
0.4\% and 1.5\% depending on the $\eta$ range considered.
The largest value is used in the full eta range.
 
Finally, the
uncertainty associated with
reweighting the simulated
events as a function of the pile-up multiplicity distribution contributes
0.6\%.
 
Table \ref{tab:EffSyst} summarises
these systematic uncertainties.
\begin{table}[h]
\centering
\caption{Summary of the uncertainties in \theRho.}
\begin{tabular}{lc}
\hline
Source & Contribution [\%] \\
\hline
Statistical                     & 0.8 \\
Kinematic reweighting (DDW)     & 0.8 \\
Muon trigger and reconstruction & 1.0 \\
BDT input variables             & 3.2 \\
Kaon tracking efficiency        & 1.5 \\
Pile-up reweighting             & 0.6 \\
\hline
\end{tabular}
\label{tab:EffSyst}
\end{table}
 
The efficiency ratio enters in Eq.~(\ref{eq:BRFormula}) with the $\mathcal{D}_{\mathrm{ref}}$ term defined in Section~\ref{sec:outline}, multiplied by the number of observed $B^\pm$ candidates. The total uncertainty in $\mathcal{D}_{\mathrm{ref}}$ is $\pm 6.3$\%.
 
A correction to the efficiency ratio for \Bsmumu  is needed because of the width difference
$\Delta \Gamma_s$ between the \Bs\ eigenstates.
According to the SM, the decay \Bsmumu\ proceeds mainly through the heavy state
$B_{s,\mathrm{H}}$~\cite{Bobeth:2013uxa,DeBruyn:2012wk}, which has width
$\Gamma_{s,\mathrm{H}}= \Gamma_s - \Delta \Gamma_s/2$, which is~$6.6$\% smaller than the
average $\Gamma_s$~\cite{PDG2018}.
The variation in the value of the \Bsmumu\ mean lifetime was tested with simulation
and found to change the \Bs\ efficiency, and consequently the
\Bs\ to \Bp\ efficiency ratio, by $+3.3$\%.
This correction is applied to the central value of $\mathcal{D}_{\mathrm{ref}}$ used
in Section~\ref{sec:br} for the determination of \BR(\Bsmumu).\footnote
{The decay time distribution of \Bsmumu\ is predicted to be different from the one of
$B_{s,\mathrm{H}}$ in scenarios of new physics, with the effect related to the observable
$\mathcal{A}^{\,\mu\mu}_{\,\Delta \Gamma}$~\cite{DeBruyn:2012wk, Buras:2013uqa,}.
The maximum possible deviation from the SM prediction of $\mathcal{A}^{\,\mu\mu}_{\,\Delta \Gamma}=+1$
is for $\mathcal{A}^{\,\mu\mu}_{\,\Delta \Gamma}=-1$,
for which the decay time distribution of \Bsmumu\ corresponds to the distribution of the
$B_{s,\mathrm{L}}$ eigenstate. In the comparison with new-physics predictions, the value of
\BR(\Bsmumu)\ obtained from this analysis should be corrected by
$+3.6\%$ or $+7.8\%$ respectively for $\mathcal{A}^{\,\mu\mu}_{\,\Delta \Gamma}=0$ and $-1$.
}
Due to the small value of $\Delta \Gamma_d$, no correction needs to be applied to
the \Bdmumu\ decay.

\section{Extraction of the signal yield}
\label{sec:sigfit}
Dimuon candidates passing the preliminary selection and the selections
against hadron misidentification and continuum background are classified according to
four intervals (with boundaries at 0.1439, 0.2455, 0.3312, 0.4163 and 1) in the BDT output.
Repeating the Run~1 analysis approach, each interval is chosen to
give an equal efficiency of $18\%$ for signal MC events,
and they are ordered according to increasing signal-to-background ratio.
 
An unbinned extended maximum-likelihood fit is performed on the dimuon invariant mass distribution
simultaneously across the four BDT intervals, each including models
for the respective signal and background contributions.
The first two bins (covering the lowest BDT values considered)
contribute mostly to background modelling: it has been
verified with MC pseudo-experiments that they have negligible impact
on the signal extraction.
 
The result of the fit is parameterised with the total yield of
\Bsmumu\ and \Bdmumu\ events in the three highest intervals of BDT output.
Section~\ref{sub:sec:model} describes
the signal and background fit models.
The parameters describing the background are allowed to vary freely and are determined by the fit.
The normalisations of the individual fit components, including the signals, are completely unconstrained and allowed to take negative values.
The ratios of the signal yields in different BDT bins are constrained
to equal the ratios of the signal efficiencies in those same
bins.
The systematic uncertainties 
due to variations in the relative signal and background efficiencies between BDT intervals,
to the signal parameterisation and to the background model are discussed in
Sections~\ref{sub:sec:BDTBins} and \ref{sub:sec:fitsyst}.
Each is modelled in the likelihood as a multiplicative Gaussian distribution whose width is equal to the corresponding systematic uncertainty.
 
\subsection{Signal and background model}
\label{sub:sec:model}
 
The signal and background models are derived from simulations and from data
collected in the mass sidebands of the search region.
 
\begin{figure}[!tbh]
\begin{center}
\includegraphics[width=0.52\textwidth]{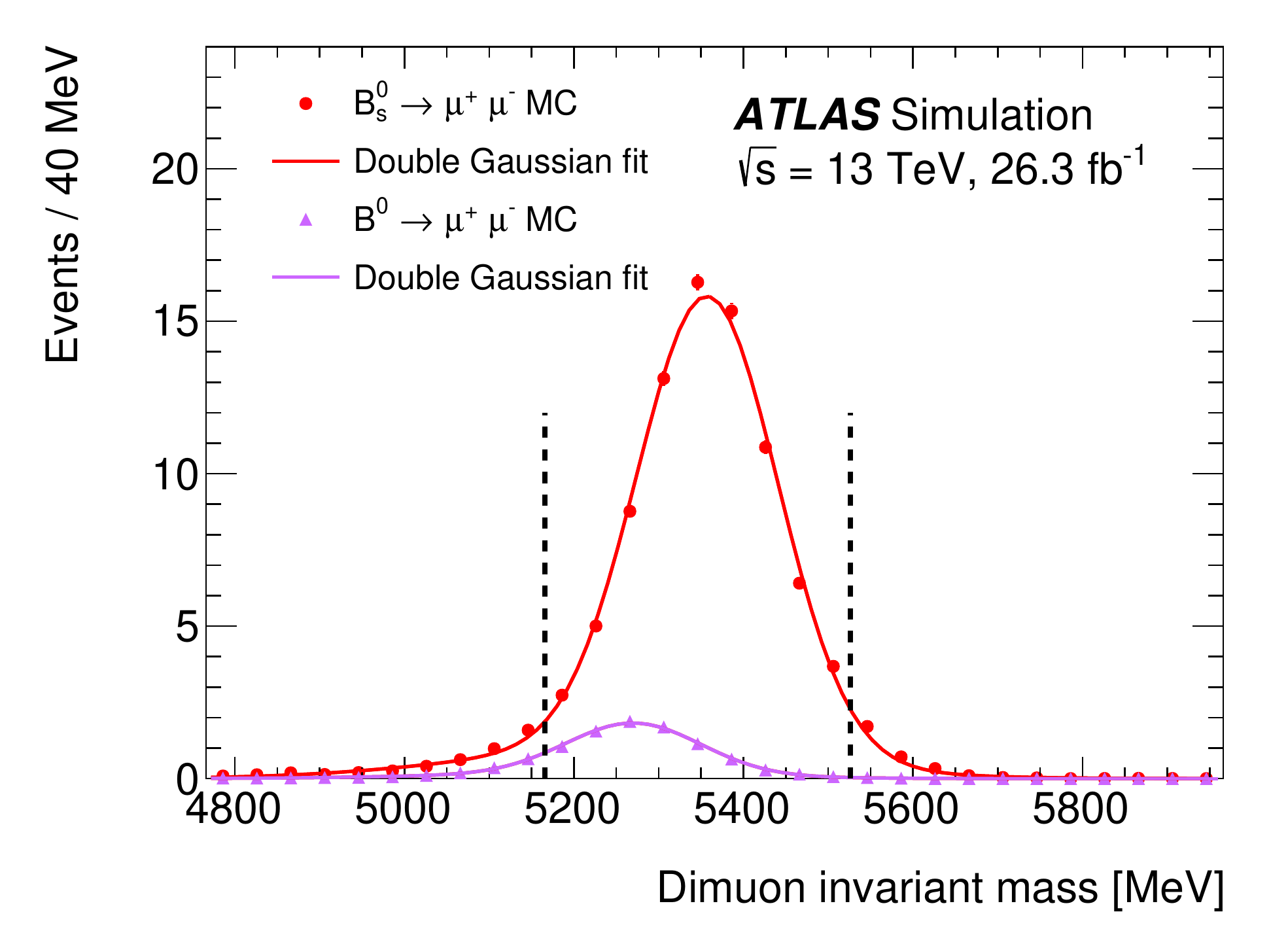}
\caption{Dimuon invariant mass distribution for the \Bs\ and \Bz\ signals
from simulation. The results of the double-Gaussian fits are overlaid.
The two distributions are normalised to the SM prediction for the expected yield with
an integrated luminosity of 26.3 fb$^{-1}$.
}
\label{fig:masssig}
\end{center}
\end{figure}

The invariant mass distribution of the \Bmumu\ signal is described with two double-Gaussian
distributions, centred respectively at the \Bz\ or \Bs\ mass.
The shape parameters are extracted from simulation, where they are found to be uncorrelated with the BDT output.
Systematic uncertainties in the mass scale and resolutions are considered separately.
Figure~\ref{fig:masssig}
shows the invariant mass distributions for \Bz\ and \Bs, obtained from MC events and normalised to
the SM expectations.
 
The background in the signal fit is composed of the types of events described in Section~\ref{sec:bkg}:
(a) the continuum background;
(b) the background from partially reconstructed $b\to \mu^+\mu^- X$ events, which is present mainly in the low mass sideband;
(c) the peaking background.
 
The non-peaking contributions have a common mass shape model, with
parameters constrained across the BDT bins in the fit as described below, and independent yields across BDT bins and components.
Systematic uncertainties arising from model assumptions will be
discussed in Section \ref{sub:sec:fitsyst}, including effects due
to the presence of $B_c^+\to J/\psi \mu^+ \nu$ and semileptonic
$B^0_{(s)}/\Lambda_b^0\to h \mu \nu$ decays.
 
Both in simulation and sideband data, the continuum background has a small linear dependence on the dimuon invariant mass. In the simulation, the slope
parameter has a roughly linear dependence versus BDT interval; the mass sidebands in data confirm this trend, albeit with large statistical uncertainty.
This dependence is included in the fit model. The small systematic uncertainties due to deviations from this assumption are discussed below in Section~\ref{sub:sec:fitsyst}.
 
The $b\to\mu^+\mu^- X$ background has a dimuon invariant mass distribution
that falls monotonically with increasing dimuon mass.
The mass dependence is derived from data in the low mass sideband, and
described with an exponential function with the same shape in each BDT interval.
The value of the shape parameter is extracted from the fit to data.
 
The invariant mass distribution of the peaking background is very similar to the \Bz\ signal,
as shown in Figure~\ref{fig:SS-SV}(b). The description of this component is obtained from MC simulation, which
indicates that the shape and normalisation are the same for all BDT bins.
In the fit, this contribution
is included with fixed mass shape and with a normalisation of $2.9 \pm 2.0$ events,
as discussed in Section~\ref{sec:fakes}.
This contribution is equally distributed among the three highest intervals of the BDT output.
 
The fitting procedure is tested with MC pseudo-experiments, as
discussed in Section \ref{sub:sec:fitsyst}.
 
\subsection{Relative signal efficiency between BDT bins}
\label{sub:sec:BDTBins}
Section \ref{sec:eff} explains how systematic uncertainties affect the overall selection
efficiency for signal candidates.
The separation of the candidates according to BDT bins introduces an additional dependence
on the relative efficiencies in each BDT bin, and systematic uncertainties in these relative
efficiencies must be accounted for.
Two different procedures are explored.
First, the distribution of the BDT output is compared between MC simulation and background-subtracted data for the reference and control channels.
The differences observed in the ratio of data to simulation are described with a linear
dependence on the BDT output. The linear dependencies observed for \BpKpJpsi\ and \BsJpsiPhi\ are in turn used to
reweight the BDT-output distribution in the \Bmumu\ MC sample.
The maximum corresponding absolute variations in the efficiencies are
equal to $+1.7$\% and $-2.3$\% respectively in the second and fourth BDT intervals, with the third interval
basically unaffected.
A second assessment of the systematic uncertainties in the relative efficiency of the BDT intervals is obtained
with a procedure similar to the one used for the event selection (Section~\ref{sec:eff}).
For each discriminating variable, the MC sample is reweighted according to the
difference between simulation and data observed in the reference channels.
The variation in the efficiency of each BDT interval is taken as the contribution
to the systematic uncertainty due to mis-modelling of that variable.
The sum in quadrature of the variations due to all discriminating variables is found to be similar in the \BpKpJpsi\ and \BsJpsiPhi\ channels.
Absolute variations of $\pm 1.0$\%, $\pm 2.4$\% and $\pm 4.4$\% are found in the second, third and fourth BDT intervals respectively.
The first of these procedures is used as a baseline for inclusion of Gaussian terms in
the signal extraction likelihood to account for the uncertainty in the relative
signal efficiency in the three most sensitive BDT bins. Care is taken
in constraining the sum of the efficiencies of the three intervals sensitive to the signal,
since that absolute efficiency and the corresponding uncertainty are parameterised with the \theRho term.
 
\begin{figure}[!t]
\begin{center}
\subfigure[]{
\hspace*{-0.5cm}
\includegraphics[width=0.52\textwidth]{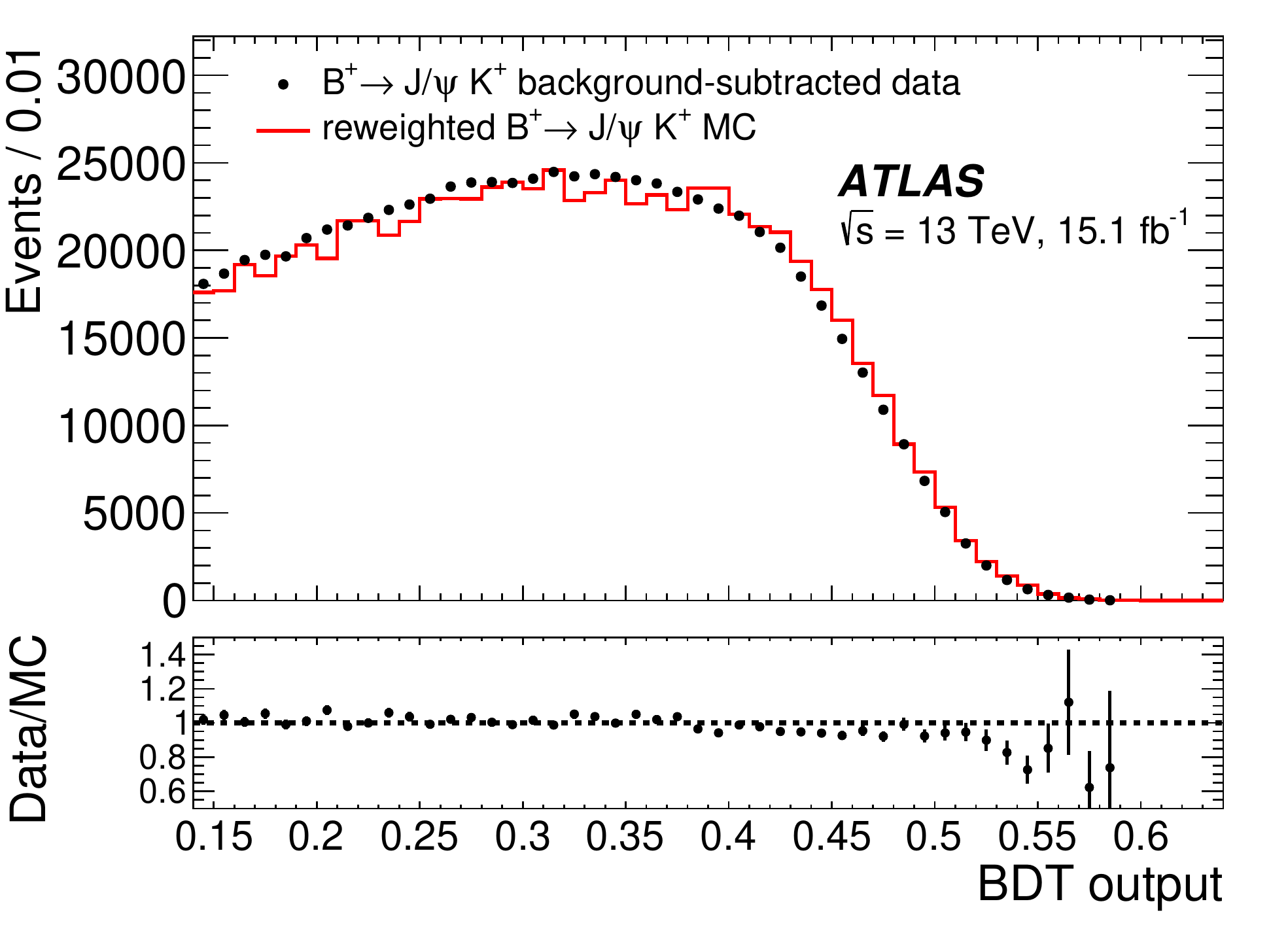}
}
\subfigure[]{
\hspace*{-0.5cm}
\includegraphics[width=0.52\textwidth]{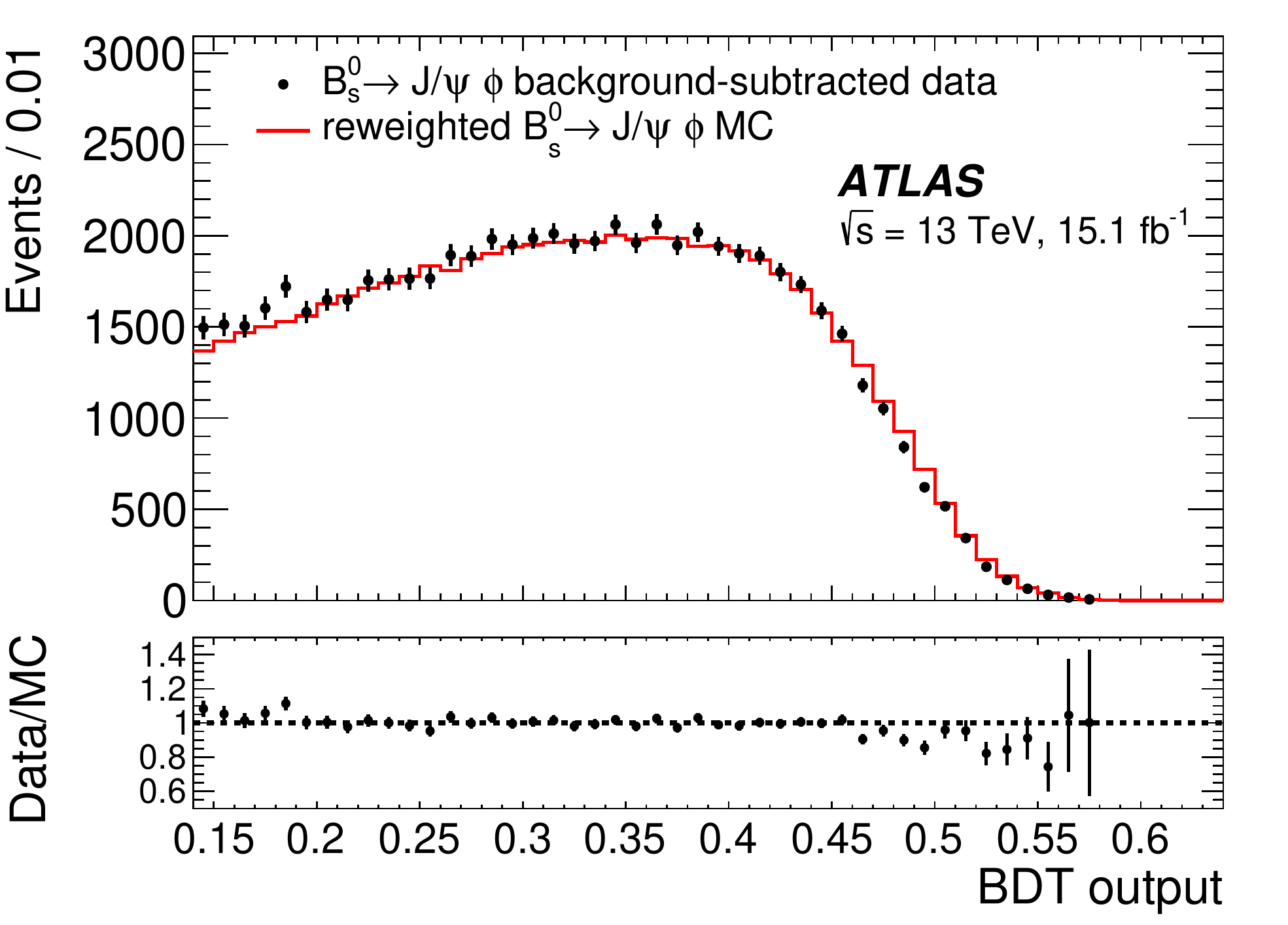}
\hspace*{-0.6cm}
}\\
\subfigure[]{
\hspace*{-0.5cm}
\includegraphics[width=0.52\textwidth]{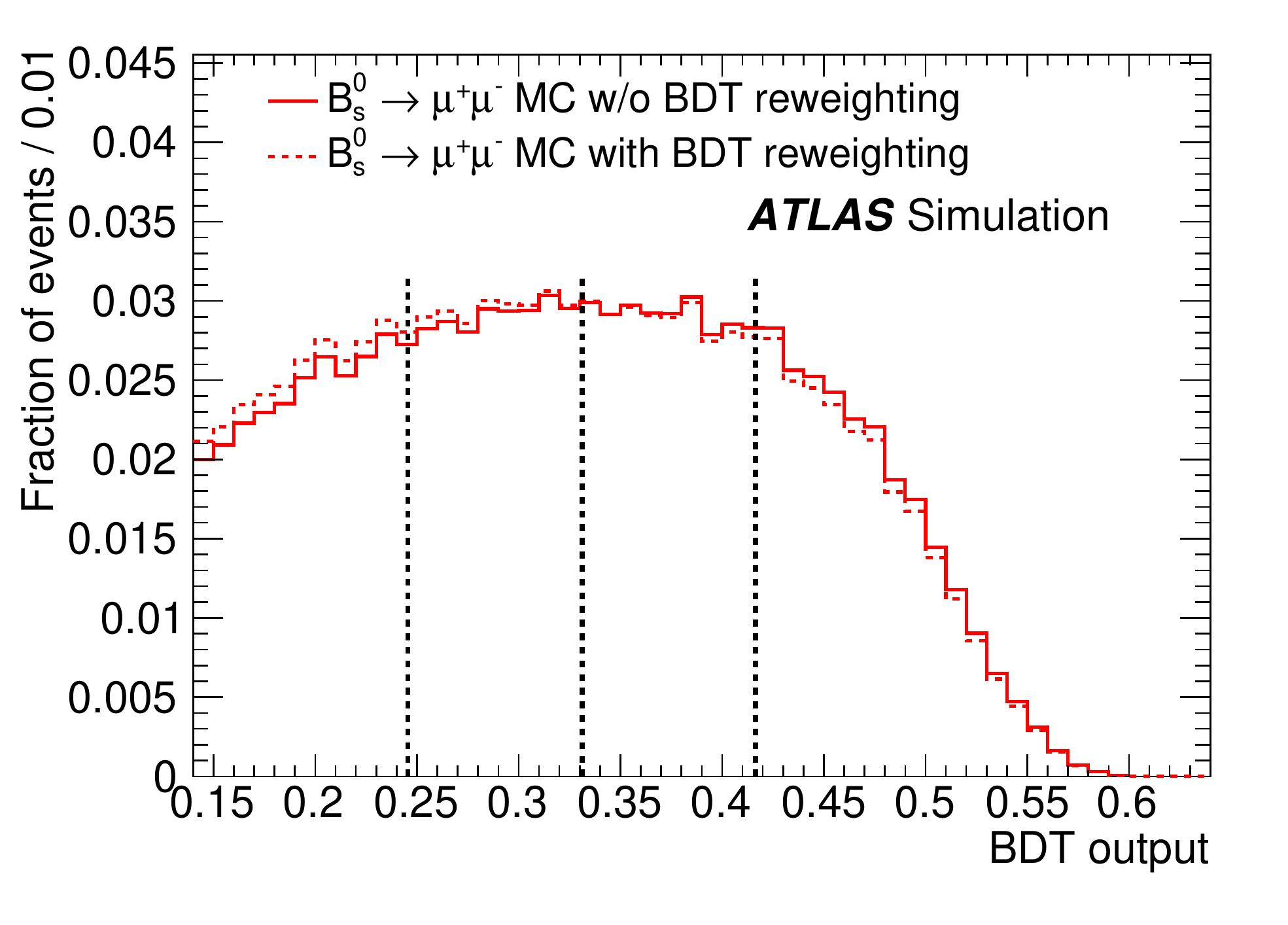}
}
\caption{BDT value distributions in data and MC simulation for (a) \BpKpJpsi, (b) \BsJpsiPhi. The MC samples are normalised to the number of data events passing
the signal reference BDT selection (Section~\ref{sec:contBDT}).
Figure (c) illustrates the BDT output for the \Bsmumu signal, with
the dashed histogram illustrating the effect of the linear reweighting on the BDT output discussed in the text. The vertical dashed lines correspond to the boundaries of the BDT intervals used in the \Bmumu\ signal fit.}
\label{fig:BDTzoomed}
\end{center}
\end{figure}

Figure~\ref{fig:BDTzoomed} shows the distribution of the BDT output from data and simulation
for the reference channels, after reweighting the MC sample.  The MC distribution for
\Bmumu\ events is also shown, illustrating how the linear deviation obtained from the reference channels
affects the simulated signal BDT output. When studying these effects, the linear fits to the ratios in Figures~\ref{fig:BDTzoomed}(a) and \ref{fig:BDTzoomed}(b)
are performed in the range corresponding to the three BDT bins with the highest signal-to-background ratio, since the remaining bin is
insensitive to the signal contribution.
 
\subsection{Systematic uncertainties in the fit}
\label{sub:sec:fitsyst}
Studies based on MC pseudo-experiments are used to assess the sensitivity
of the fit to the input assumptions.
Variations in the description of signal and background components are used
in the generation of these samples.
The corresponding changes in the average
numbers, $N_s$ and $N_d$, of \Bs\ and \Bz\ events determined by the fit,
run in the nominal configuration, are taken as systematic uncertainties.
The size of the variations used in the generation of the MC pseudo-experiments
is determined in some cases by known characteristics of the ATLAS detector
(reconstructed momentum scale and momentum resolution), in others using MC evaluation
(background due to semileptonic $B^0_{(s)}/\Lambda_b^0\to h \mu \nu$ decays and to
$B_c^+ \to J/\psi \mu^+\nu$),
and in others from uncertainties determined from data in the sidebands
or from simulation (shapes of the background components and their variation across
the BDT intervals).
 
The MC pseudo-experiments were generated
with the normalisation of the continuum and $b\to\mu\mu X$ components obtained from the fit
to the data in the sidebands of the invariant mass distribution,
and the peaking background from the expectation discussed in Section~\ref{sec:fakes}.
The signal was generated with different configurations, roughly covering the
range between zero and twice the expected SM yield.
 
For all variations of the assumptions and all configurations of the signal amplitudes
the distributions of the differences between fit results and generated values are used to evaluate systematic uncertainties.
In addition, distributions obtained from MC pseudo-experiments generated and fitted according to the nominal fit model
are used to study systematic biases deriving from the fit procedure.
For both signal yields, the bias is smaller than 15\% of the fit error, for true values of the \Bsmumu\ branching fraction above $5\times 10^{-10}$.
 
The shifts in $N_s$ or $N_d$ are combined by considering separately
the sums in quadrature of the positive and negative shifts
and taking the larger as the symmetric systematic uncertainty.
The total systematic uncertainty is found to increase with the assumed size
of the signal, with a dependence
$\sigma_\mathrm{syst}^{N_s} = 3 + 0.05 N_s$ and $\sigma_\mathrm{syst}^{N_d} = 2.9 + 0.05 N_s + 0.05 N_d$.
Most of the shifts observed have opposite sign for $N_s$ and $N_d$, resulting
in a combined correlation coefficient in the systematic uncertainties
of $\rho_\mathrm {syst}=-0.83$.
 
The systematic uncertainties discussed in this Section are included in the
fit to the $\mumu$ candidates in data.
The fit for the yield of \Bs\ and \Bz\ events is modified by including in the likelihood
two smearing parameters for $N_s$ and $N_d$
that are constrained by a two-dimensional Gaussian distribution parameterised by the values of
$\sigma_\mathrm{syst}^{N_s}$, $\sigma_\mathrm {syst}^{N_d}$ and  $\rho_\mathrm {syst}$.
 
\begin{figure}[!t]
\begin{center}
\subfigure[]{
\includegraphics[width=0.48\textwidth]{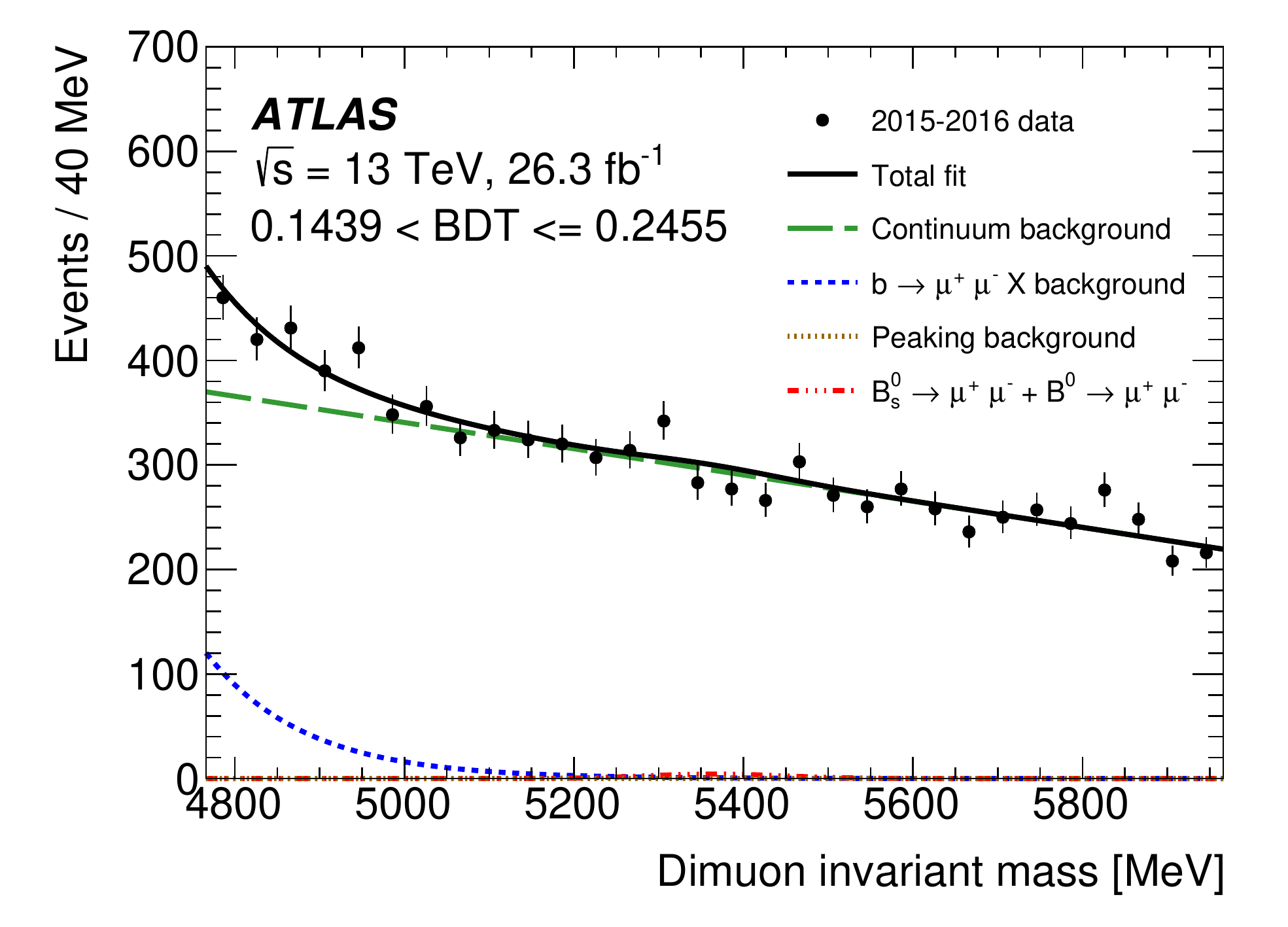}
}
\subfigure[]{
\includegraphics[width=0.48\textwidth]{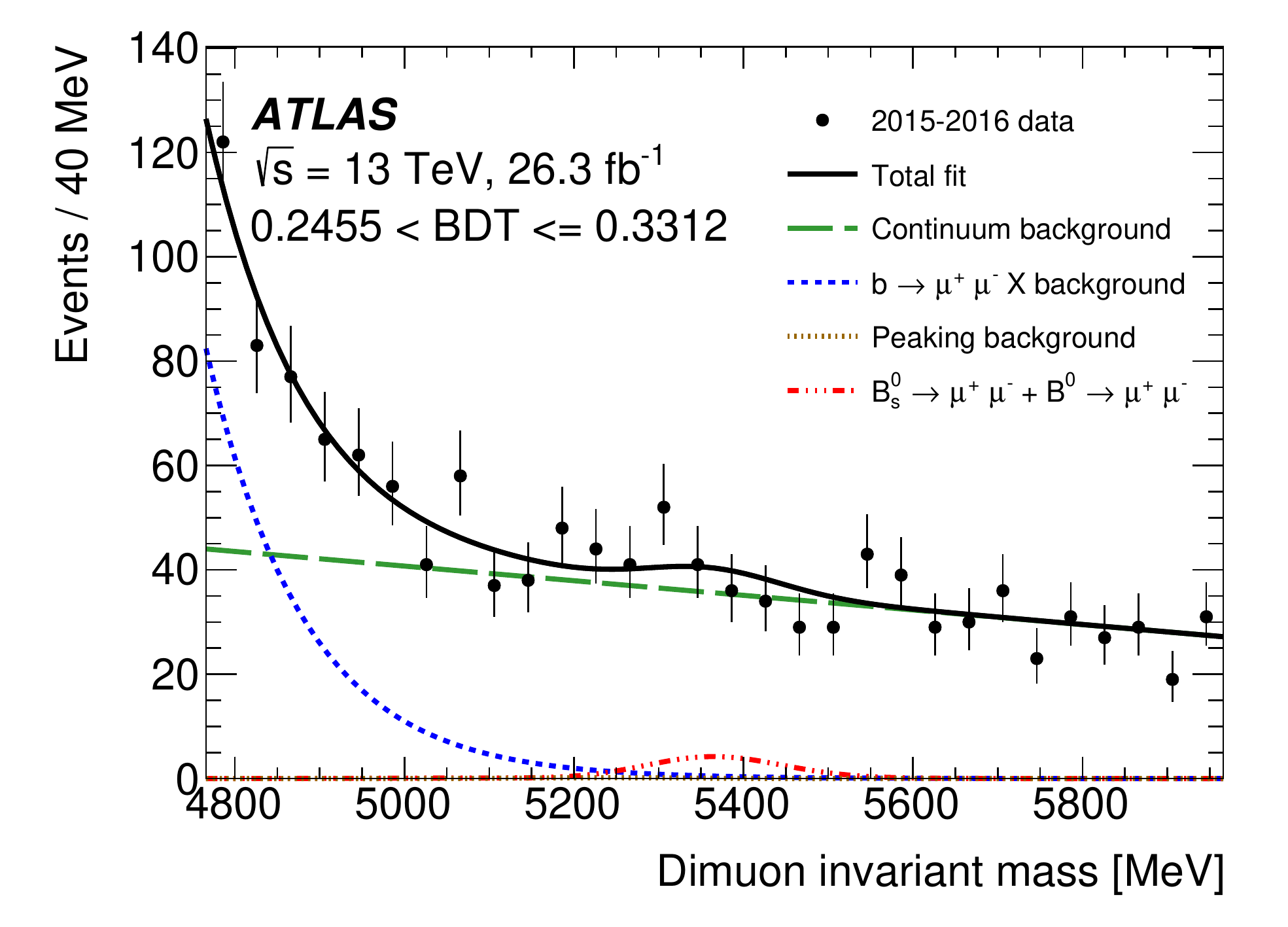}
}\\
\subfigure[]{
\includegraphics[width=0.48\textwidth]{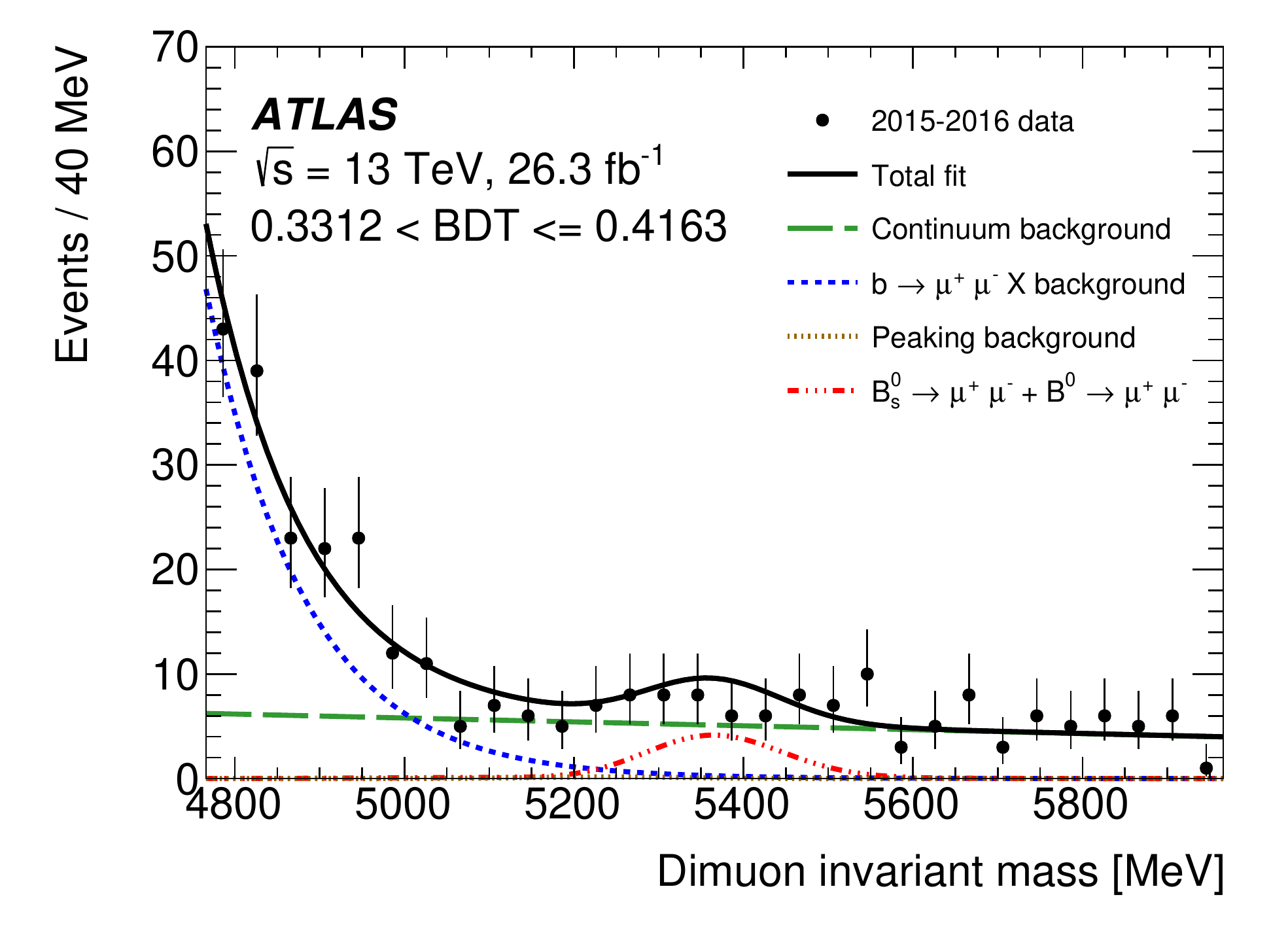}
}
\subfigure[]{
\includegraphics[width=0.48\textwidth]{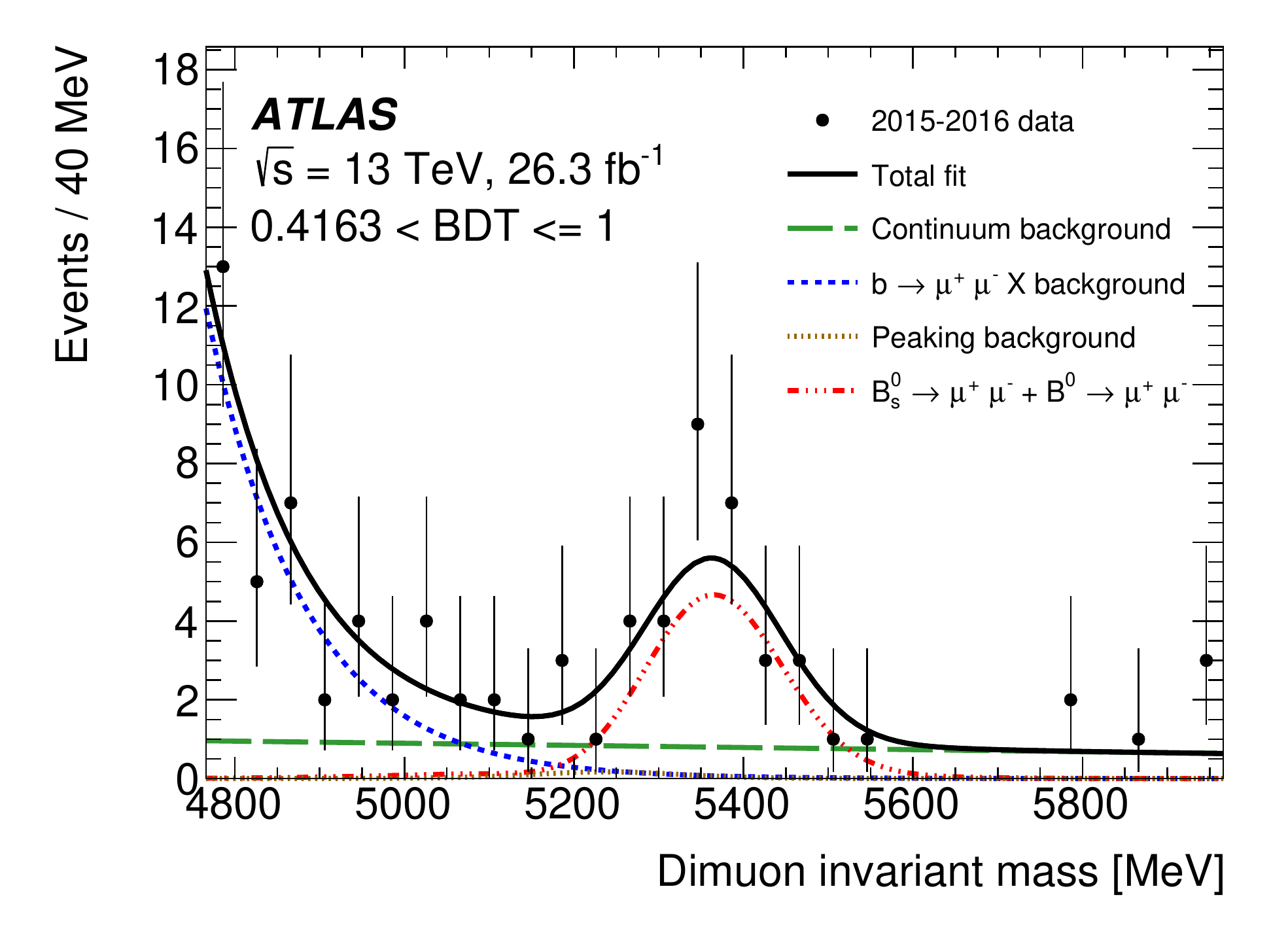}
}
\caption{
Dimuon invariant mass distributions in the unblinded data,
in the four intervals of BDT output. Superimposed is the
result of the maximum-likelihood fit.  The total fit is
shown as a continuous line, with the dashed lines corresponding
to the observed signal component, the $b\to\mu\mu X$ background,
and the continuum background. The signal components are grouped
in one single curve, including both the \Bsmumu\ and the (negative) \Bdmumu\ component.
The curve representing the peaking \Bhh background lies very close to the horizontal axis in all BDT bins.}
\label{fig:unblinded_default}
\end{center}
\end{figure}
 
\subsection{Results of the signal yield extraction}
 
The numbers of background events contained in the signal region ($5166$--$5526$~\MeV)
are computed from the interpolation of the data observed in the sidebands.
This procedure yields $2685 \pm 37$, $330 \pm 14$, $51 \pm 6$ and $7.9 \pm 2.6$ events respectively
in the four intervals of BDT output. For comparison, the total expected numbers of
signal events according to the SM prediction are $91$ and $10$ for $N_s$ and $N_d$ respectively,
equally distributed among the three intervals with
the highest signal-to-background ratio.
 
In those three BDT intervals, in the unblinded signal region, a total of $1951$ events in the full mass range of
$4766$--$5966$~MeV are used in the likelihood fit to signal and background.
Without applying any bounds on the values of the fitted parameters,
the values determined by the fit are
$N_s = 80 \pm 22$ and $N_d = - 12 \pm 20$,
where the uncertainties correspond to likelihood variations satisfying $- 2\,\Delta \ln (\mathcal{L}) = 1$.
The likelihood includes the systematic uncertainties discussed above, but
statistical uncertainties largely dominate. The result is consistent with the expectation from simulation.
The uncertainties in the result of the fit are discussed in Section~\ref{sec:br},
where the measured values of the branching fractions are presented.
 
Figure~\ref{fig:unblinded_default}
shows the dimuon invariant mass distributions
in the four BDT intervals, together with
the projections of the likelihood. A modified Kolmogorov--Smirnov (KS) test~\cite{Kolmogorov} is used to estimate the fit quality:
the $p$-value is estimated by comparing the maximum of the KS distance across the four histograms of Figure~\ref{fig:unblinded_default} with
the distribution of the same quantity from pseudo-experiments generated with the shape resulting from the fit to data. This procedure yields a compatibility probability of 84\%.

\section{Branching fraction extraction}
\label{sec:br}
 
The branching fractions for the decays \Bsmumu\ and \Bdmumu\  are extracted from data
using a maximum-likelihood fit.  The likelihood is obtained from the one
used for $N_s$ and $N_d$ by replacing the fit parameters
with the corresponding branching fractions divided by normalisation terms
in Eq.~(\ref{eq:BRFormula}), and including Gaussian multiplicative factors
for the normalisation uncertainties.
All results are obtained profiling the fit likelihood with respect to all parameters involved
other than the branching fraction(s) of interest.
 
The normalisation terms include external inputs for the $\Bp$ branching fraction
and the relative hadronisation probability.
The branching fraction is obtained from world averages~\cite{PDG2018} as the product of
$\BR(\BpKpJpsi)=(1.010 \pm 0.029) \times 10^{-3}$
and $\BR(\Jpsi\to\mumu)=(5.961 \pm 0.033)\%$.
The relative hadronisation probability is equal to one for $\Bz$, while for $\Bs$ it is taken from the latest available
HFLAV average~\cite{HFLAV2016} $f_s/f_d = 0.256 \pm 0.013$, which assumes $f_u/f_d=1$.
 
The efficiency-weighted and luminosity-weighted number of events for the reference
channel enters in Eq.~(\ref{eq:BRFormula}) with the denominator $\mathcal{D}_{\mathrm{ref}}$.
The values $\mathcal{D}_{\mathrm{ref}} =(5.69 \pm 0.36) \times 10^{6}$ for $\Bs$
and $(5.84 \pm 0.37) \times 10^{6}$ for $\Bz$ are obtained from the results of
Sections~\ref{sec:bplus} and~\ref{sec:eff},
and include the correction to the \Bsmumu\ efficiency due to the
lifetime difference between $B_{s,\mathrm{H}}$ and $\Bs$.
The combination of $\Bp$ branching fraction, hadronisation probabilities and
$\mathcal{D}_{\mathrm{ref}}$, i.e. the single-event sensitivity,
is equal to
$(4.02\pm 0.35)\times 10^{-11}$ for
\Bsmumu\ and $(1.059\pm0.074)\times 10^{-11}$ for \Bdmumu.
 
\begin{figure}[!t]
\begin{center}
\subfigure[]{
\hspace*{-0.5cm}
\includegraphics[width=0.49\textwidth]{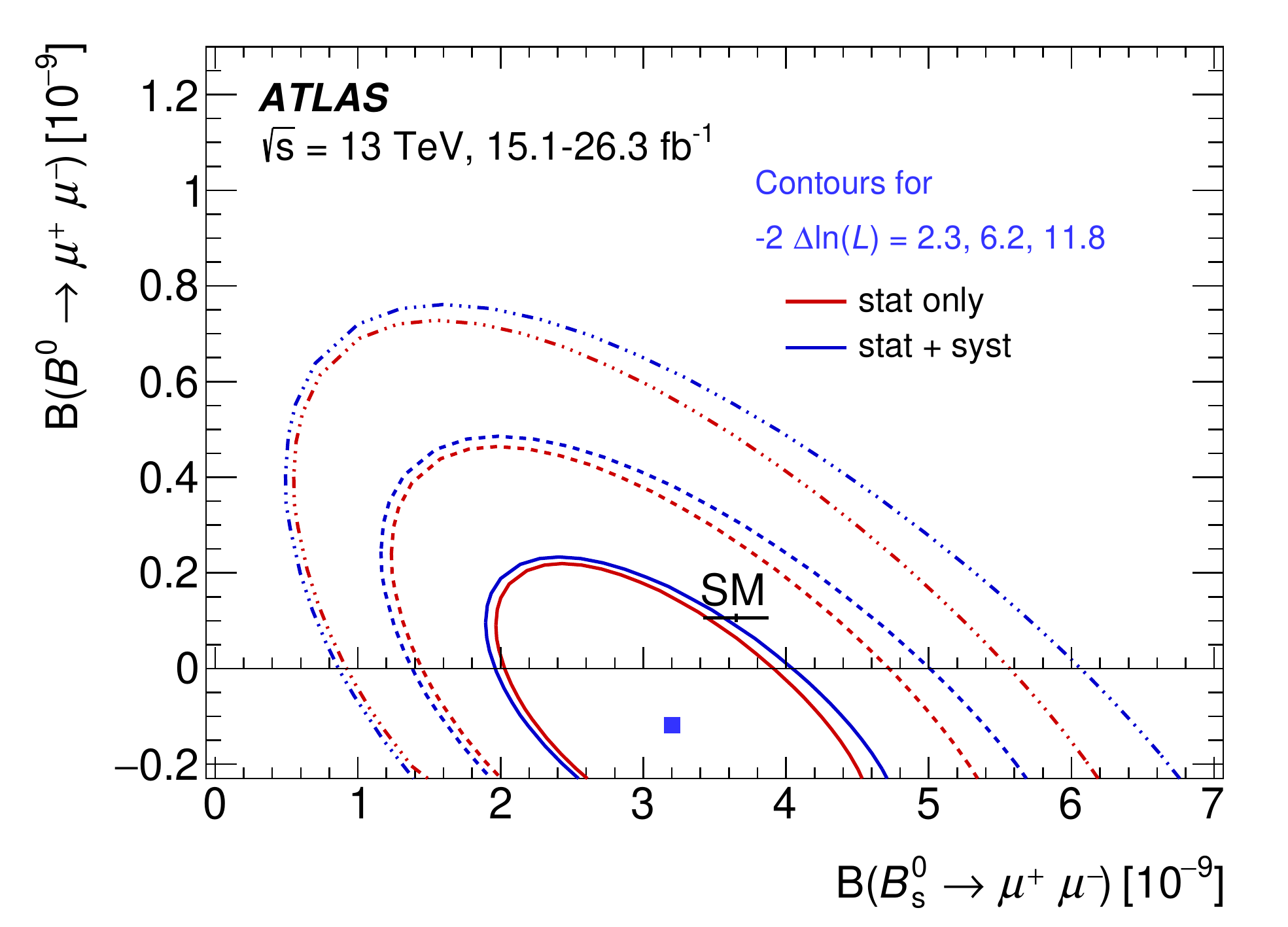}
}
\subfigure[]{
\hspace*{-0.5cm}
\includegraphics[width=0.49\textwidth]{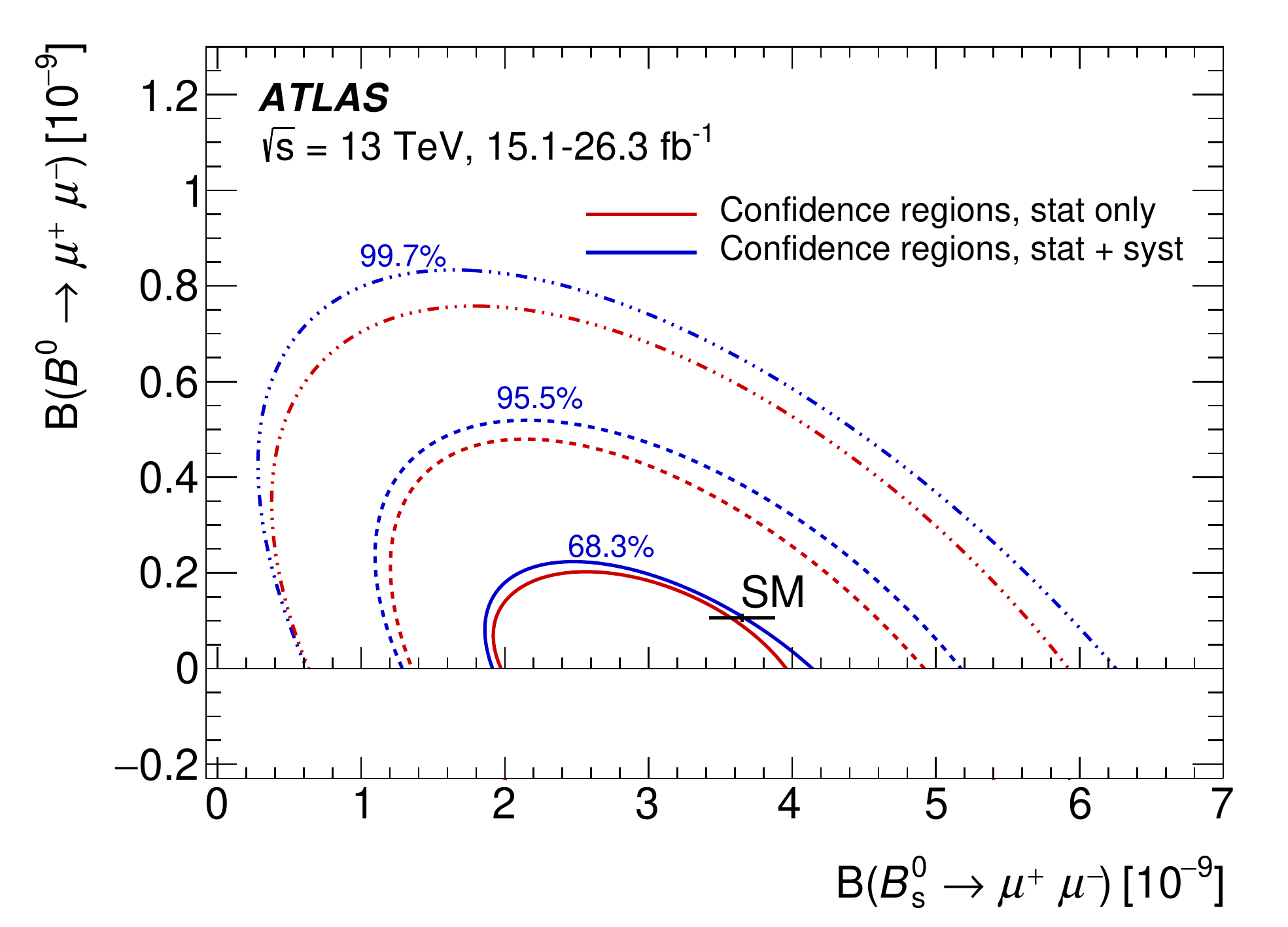}
}
\caption
{(a)
Likelihood contours for the simultaneous fit to \BR(\Bsmumu)\ and \BR(\Bdmumu),
for values of $-2\Delta \ln\left(\mathcal{L}\right)$ equal to 2.3, 6.2 and 11.8.
The SM prediction with uncertainties is indicated.
(b) Neyman contours in the $\BR (\Bsmumu)$--$\BR (\Bdmumu)$ plane for 68.3\%, 95.5\% and 99.7\% coverage.
At each $-2\Delta \ln\left(\mathcal{L}\right)$ or coverage value, the inner contours are statistical uncertainty only,
while the outer ones include statistical and systematic uncertainties. The construction of these contours makes
use of both the dimuon (26.3 \ifb) and the reference channel (15.1 \ifb) datasets.
}
\label{fig:like-contours-unconstrained}
\end{center}
\end{figure}

The values of the branching fractions that maximise the profile likelihood 
are $\BR (\Bsmumu) =  \left( 3.2 \pm 0.9 \right) \times 10^{-9}$ and $\BR (\Bdmumu) = \left( -1.3 \pm 2.1\right)\times 10^{-10}$,
where the uncertainties, both statistical and systematic combined, are estimated using a
Gaussian approximation in the neighbourhood of the likelihood function maximum.
Figure~\ref{fig:like-contours-unconstrained}(a) shows likelihood contours for the simultaneous fit to \BR(\Bsmumu)\ and \BR(\Bdmumu),
for values of $-2\Delta \ln\left(\mathcal{L}\right)$ equal to 2.3, 6.2 and 11.8, relative to the maximum of the likelihood.
 
Table~\ref{tab:BRSyst} gives a breakdown of the estimated contributions of systematic and statistical uncertainties.
The results are dominated by statistical uncertainties, with the most prominent source of systematic uncertainty coming from fit
uncertainties, where the largest contributors are the
mass scale and $b\to \mumu X$ background parameterisation.
 
\begin{table}
\centering
\caption{Breakdown of the expected systematic uncertainties in \BR(\Bmumu). The measurements are dominated
by statistical uncertainty, followed by the systematic uncertainty from the fit. The latter is dominated by contributions from the mass scale uncertainty
and the parameterisation of the $b\to \mumu X$ background. The statistical uncertainties reported here are obtained from the maximisation
of the fit likelihood and are meant only as a reference for the relative scale uncertainties.}
\begin{tabular}{lcc}
\hline
Source & \Bs [\%] & \Bz [\%] \\
\hline
$f_s/f_d$                                   & 5.1  & -    \\
$B^+$ yield                                 & 4.8  & 4.8  \\
$\theRho$                                   & 4.1  & 4.1  \\
$\BR(\BpKpJpsi)\times\BR(\Jpsi\to\mumu)$    & 2.9  & 2.9  \\
Fit systematic uncertainties                & 8.7  & 65 \\
\hline
Stat. uncertainty (from likelihood est.)    & 27 & 150  \\
\hline
\end{tabular}
\label{tab:BRSyst}
\end{table}
 
Given the statistical regime of the analysis, the likelihood contours of
Figure~\ref{fig:like-contours-unconstrained}(a) cannot be
immediately translated into contours with the conventional coverage of one, two and three Gaussian standard deviations. Moreover,
the contours extend into regions of negative branching fractions, which are unphysical.
In order to address these points, a Neyman construction~\cite{neyman} is employed to obtain the 68.3\%, 95.5\% and 99.7\%
confidence intervals in the \BR(\Bsmumu)\,--\,\BR(\Bdmumu) plane. This construction
yields the contours shown in Figure~\ref{fig:like-contours-unconstrained}(b).
The same construction is used to determine the $68.3\%$ confidence interval
for \BR(\Bsmumu) with MC pseudo-experiments, and results in
\begin{equation}
\BR (\Bsmumu) =  \left( 3.21^{+0.96 +0.49}_{-0.91 -0.30} \right) \times 10^{-9}=  \left( 3.2^{+1.1}_{-1.0} \right) \times 10^{-9}\, .\notag
\end{equation}
Statistical and systematic uncertainties (shown separately in the first expression, and combined in the last)
are separated by repeating the likelihood fit after setting all systematic uncertainties to zero.
The fitted branching fractions (obtained from the unconstrained likelihood maximum) are in all cases inputs to the Neyman construction,
which, by design, results in physically allowed values for the resulting branching fractions.
 
The upper limit at 95\% CL on \BR(\Bdmumu), determined with the same Neyman procedure, is
\begin{equation}
\BR (\Bdmumu) <  4.3~\times~10^{-10}\, .\notag
\end{equation}
 
Using the predicted SM branching fractions from Section~\ref{sec:introduction},
the analysis is expected to yield on average a measurement of
$\left( 3.6^{+1.1}_{-1.0} \right)~\times~10^{-9}$
for \BR (\Bsmumu) and an upper limit at 95\% CL of $7.1~\times~10^{-10}$ for \BR (\Bdmumu).

\section{Combination with the Run~1 result}
\label{sec:combination}
The likelihood function from the current result is combined with the likelihood function from the Run~1 result~\cite{Aaboud:2016ire}.
The only common parameters in the combination are the fitted \BR(\Bmumu) and the combination of external inputs
$\mathcal{F}_{\mathrm{ext}} = \BR(\BpKpJpsi)\times\BR(\Jpsi\to\mumu)\times(f_{u}/f_{s}) = (2.35 \pm 0.14) \times 10^{-4}$.
Except for $\mathcal{F}_{\mathrm{ext}}$, all nuisance parameters are treated as uncorrelated between the two likelihoods, with
both likelihoods including their individual parameterisations of systematic uncertainty effects.
A negligible change in the results,
corresponding to shifts in central values and uncertainties between 1\% and 4\%, is
found when all sources of systematic uncertainty are assumed to be fully correlated.
 
The maximum of the combined likelihood
corresponds to
\begin{eqnarray*}
\BR (\Bsmumu) =  \left(  2.8  \pm 0.7 \right) \times 10^{-9}\, ,\notag \\
\BR (\Bdmumu) =  \left( -1.9  \pm 1.6  \right) \times 10^{-10}\, .\notag
\end{eqnarray*}
 
\begin{figure}[h]
\begin{center}
\hspace{-1.2cm}
\includegraphics[width=0.98\textwidth]{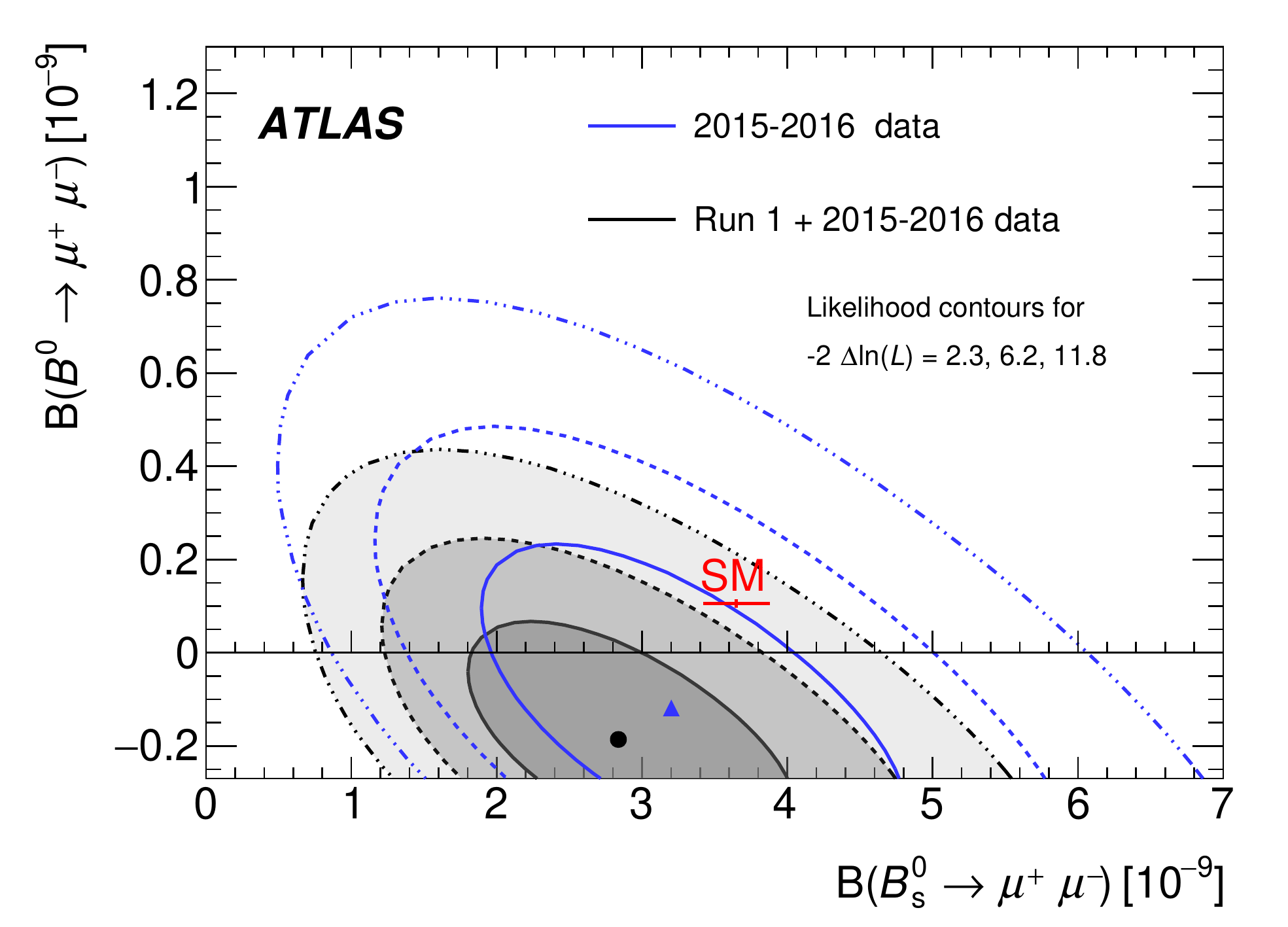}
\caption
{Likelihood contours for the combination of the Run~1 and 2015--2016 Run~2 results (shaded areas).
The contours are obtained
from the combined likelihoods of the two analyses,
for values of $-2\Delta \ln\left(\mathcal{L}\right)$ equal to 2.3, 6.2 and 11.8.
The empty contours represent the result from 2015--2016 Run~2 data alone. The SM prediction with uncertainties is indicated.
}
\label{app:fig:2DLH-run1comb}
\end{center}
\end{figure}

Figure~\ref{app:fig:2DLH-run1comb} shows the likelihood contours for the
combined Run~1 and Run~2 result for
\BR(\Bsmumu)\ and \BR(\Bdmumu), for values of $-2\Delta \ln\left(\mathcal{L}\right)$ equal to 2.3, 6.2 and 11.8, relative to the maximum of the likelihood. The contours for the result from 2015--2016 Run~2 data are overlaid for comparison.
 
When applying the one-dimensional Neyman construction described in Section \ref{sec:br} to this combined likelihood, whose maximum is unconstrained and allowed to access the unphysical (negative) region,
the $68.3\%$ confidence interval obtained
for \BR(\Bsmumu) is
\begin{equation}
\BR (\Bsmumu) =  
\left( 2.8^{+0.8}_{-0.7} \right) \times 10^{-9}\, .\notag
\end{equation}
The upper limit at 95\% CL on \BR(\Bdmumu) is determined with the same Neyman procedure, yielding
\begin{equation}
\BR (\Bdmumu) <  2.1~\times~10^{-10}\, .\notag
\end{equation}
 
Using the predicted SM branching fractions from Section~\ref{sec:introduction},
the analysis is expected to yield on average a measurement of
$\left( 3.6^{+0.9}_{-0.8} \right)~\times~10^{-9}$
for \BR (\Bsmumu) and an upper limit of $5.6~\times~10^{-10}$ for \BR (\Bdmumu).
 
The Run~1 and Run~2 results are found to be 1.2 standard deviations apart.
Using both runs,
the combined significance of the \Bsmumu\ signal is
estimated to be 4.6 standard deviations,
and the combined branching fraction measurements differ by
2.4 standard deviations from the SM values in the \BR (\Bdmumu)--\BR (\Bsmumu) plane. These significancies are
assessed purely from the evaluation of likelihood ratios.

\section{Conclusions}
 
A study of the rare decays of \Bs and \Bz mesons into oppositely charged muon pairs is presented,
based on $36.2$~fb$^{-1}$ of $13$ \TeV\ LHC proton--proton collision data
collected by the ATLAS experiment in 2015 and 2016.

For the \Bs the branching fraction is determined to be $\BR (\Bsmumu) = \left( 3.2^{+1.1}_{-1.0} \right) \times 10^{-9}$,
where the uncertainty includes both the statistical and systematic contributions. The result is consistent with the analysis expectation
of $\left( 3.6^{+1.1}_{-1.0}\right)~\times~10^{-9}$ in the SM hypothesis.
For the \Bz an upper limit $\BR (\Bdmumu) < 4.3 \times 10^{-10}$ is placed at the 95\% confidence level,
with an expected upper bound of $7.1~\times~10^{-10}$ in the SM hypothesis.
The limit is compatible with the SM prediction.

The result presented in this paper is combined with the ATLAS result from the full Run~1 dataset to obtain
$\BR (\Bsmumu) =  \left(2.8^{+0.8}_{-0.7}\right) \times 10^{-9}$ and $\BR (\Bdmumu) < 2.1 \times 10^{-10}$.
All the results presented are compatible with the branching fractions predicted by the SM as well as with currently available experimental results.
 
\section*{Acknowledgements}
 
 
We thank CERN for the very successful operation of the LHC, as well as the
support staff from our institutions without whom ATLAS could not be
operated efficiently.
 
We acknowledge the support of ANPCyT, Argentina; YerPhI, Armenia; ARC, Australia; BMWFW and FWF, Austria; ANAS, Azerbaijan; SSTC, Belarus; CNPq and FAPESP, Brazil; NSERC, NRC and CFI, Canada; CERN; CONICYT, Chile; CAS, MOST and NSFC, China; COLCIENCIAS, Colombia; MSMT CR, MPO CR and VSC CR, Czech Republic; DNRF and DNSRC, Denmark; IN2P3-CNRS, CEA-DRF/IRFU, France; SRNSFG, Georgia; BMBF, HGF, and MPG, Germany; GSRT, Greece; RGC, Hong Kong SAR, China; ISF and Benoziyo Center, Israel; INFN, Italy; MEXT and JSPS, Japan; CNRST, Morocco; NWO, Netherlands; RCN, Norway; MNiSW and NCN, Poland; FCT, Portugal; MNE/IFA, Romania; MES of Russia and NRC KI, Russian Federation; JINR; MESTD, Serbia; MSSR, Slovakia; ARRS and MIZ\v{S}, Slovenia; DST/NRF, South Africa; MINECO, Spain; SRC and Wallenberg Foundation, Sweden; SERI, SNSF and Cantons of Bern and Geneva, Switzerland; MOST, Taiwan; TAEK, Turkey; STFC, United Kingdom; DOE and NSF, United States of America. In addition, individual groups and members have received support from BCKDF, CANARIE, CRC and Compute Canada, Canada; COST, ERC, ERDF, Horizon 2020, and Marie Sk{\l}odowska-Curie Actions, European Union; Investissements d' Avenir Labex and Idex, ANR, France; DFG and AvH Foundation, Germany; Herakleitos, Thales and Aristeia programmes co-financed by EU-ESF and the Greek NSRF, Greece; BSF-NSF and GIF, Israel; CERCA Programme Generalitat de Catalunya, Spain; The Royal Society and Leverhulme Trust, United Kingdom.
 
The crucial computing support from all WLCG partners is acknowledged gratefully, in particular from CERN, the ATLAS Tier-1 facilities at TRIUMF (Canada), NDGF (Denmark, Norway, Sweden), CC-IN2P3 (France), KIT/GridKA (Germany), INFN-CNAF (Italy), NL-T1 (Netherlands), PIC (Spain), ASGC (Taiwan), RAL (UK) and BNL (USA), the Tier-2 facilities worldwide and large non-WLCG resource providers. Major contributors of computing resources are listed in Ref.~\cite{ATL-GEN-PUB-2016-002}.
 

\printbibliography
 
\clearpage
 
\begin{flushleft}
{\Large The ATLAS Collaboration}

\bigskip

M.~Aaboud$^\textrm{\scriptsize 34d}$,    
G.~Aad$^\textrm{\scriptsize 99}$,    
B.~Abbott$^\textrm{\scriptsize 125}$,    
D.C.~Abbott$^\textrm{\scriptsize 100}$,    
O.~Abdinov$^\textrm{\scriptsize 13,*}$,    
B.~Abeloos$^\textrm{\scriptsize 129}$,    
D.K.~Abhayasinghe$^\textrm{\scriptsize 91}$,    
S.H.~Abidi$^\textrm{\scriptsize 164}$,    
O.S.~AbouZeid$^\textrm{\scriptsize 39}$,    
N.L.~Abraham$^\textrm{\scriptsize 153}$
% ...
\bigskip
\\

% ...
$^{13}$Institute of Physics, Azerbaijan Academy of Sciences, Baku; Azerbaijan.\\
% ...
$^{34}$$^{(a)}$Facult\'e des Sciences Ain Chock, R\'eseau Universitaire de Physique des Hautes Energies - Universit\'e Hassan II, Casablanca;$^{(b)}$Centre National de l'Energie des Sciences Techniques Nucleaires (CNESTEN), Rabat;$^{(c)}$Facult\'e des Sciences Semlalia, Universit\'e Cadi Ayyad, LPHEA-Marrakech;$^{(d)}$Facult\'e des Sciences, Universit\'e Mohamed Premier and LPTPM, Oujda;$^{(e)}$Facult\'e des sciences, Universit\'e Mohammed V, Rabat; Morocco.\\
% ...
$^{39}$Niels Bohr Institute, University of Copenhagen, Copenhagen; Denmark.\\
% ...
$^{91}$Department of Physics, Royal Holloway University of London, Egham; United Kingdom.\\
% ...
$^{99}$CPPM, Aix-Marseille Universit\'e, CNRS/IN2P3, Marseille; France.\\
$^{100}$Department of Physics, University of Massachusetts, Amherst MA; United States of America.\\
% ...
$^{125}$Homer L. Dodge Department of Physics and Astronomy, University of Oklahoma, Norman OK; United States of America.\\
% ...
$^{129}$LAL, Universit\'e Paris-Sud, CNRS/IN2P3, Universit\'e Paris-Saclay, Orsay; France.\\
% ...
$^{153}$Department of Physics and Astronomy, University of Sussex, Brighton; United Kingdom.\\
% ...
$^{164}$Department of Physics, University of Toronto, Toronto ON; Canada.\\
% ...
$^{a}$ Also at Borough of Manhattan Community College, City University of New York, NY; United States of America.\\
$^{b}$ Also at California State University, East Bay; United States of America.\\
$^{c}$ Also at Centre for High Performance Computing, CSIR Campus, Rosebank, Cape Town; South Africa.\\
$^{d}$ Also at CERN, Geneva; Switzerland.\\
$^{e}$ Also at CPPM, Aix-Marseille Universit\'e, CNRS/IN2P3, Marseille; France.\\
% ...
$^{*}$ Deceased

\end{flushleft}


\end{document}